# GENETIC CODE: FOUR DIVERSITY TYPES OF PROTEIN AMINO ACIDS

**Miloje M. Rakočević***

Department of Chemistry, Faculty of Science, University of Niš, Serbia

## Abstract

This paper presents, for the first time, four diversity types of protein amino acids. The first type includes two amino acids (G, P), both without standard hydrocarbon side chains; the second one four amino acids, as two pairs [(A, L), (V, I)], all with standard hydrocarbon side chains; the third type comprises the six amino acids, as three pairs [(F, Y), (H, W), (C, M)], two aromatic, two hetero aromatic and two "hetero" non-aromatic); finally, the fourth type consists of eight amino acids, as four pairs [(S, T), (D, E), (N, Q), (K, R)], all with a functional group which also exists in amino acid functional group (wholly presented: $H_2N$-ĊH-COOH; separately: OH, COOH, $CONH_2$, $NH_2$). The insight into existence of four types of diversity was possible only after an insight into the existence of some very new arithmetical regularities, which were so far unknown. Also, as for showing these four types was necessary to reveal the relationships between several key harmonic structures of the genetic code (which we presented in our previous works), this paper is also a review article of the author's researches of the genetic code. By this, the review itself shows that the said harmonic structures are connected through the same (or near the same) chemically determined amino acid pairs, 10 pairs out of the 190 possible.

---

* Now retired, on the address: Milutina Milankovica 118/ 25, 11070 Belgrade, Serbia; e-mail: milemirkov@open.telekom.rs; website: (http://www.rakocevcode.rs).

## 1. INTRODUCTION

Eighteen years ago V. Shcherbak pointed out that the classifications and systematizations of the genetic code constituents (protein, i.e. canonical amino acids, AAs) "appears to be loaded with elegant arithmetical regularities, which are so far unknown", and also that "the physical nature of such a phenomenon is so far not clear" (Shcherbak, 1993).

By this, the mentioned classifications-systematizations and the arithmetical regularities in which "the amino acids appear to be divided into two parts" are realized through the number of nucleons just in these two parts (Shcherbak, 1993, 1994). Surprisingly, since then no attempt has been made to answer two questions which follow from Shcherbak's finding: 1. how to explain such regularities, which had never before been known to mathematics; and in particular, how to explain their presence in the genetic code; and 2. what is the physical (and chemical) nature of the presence of these regularities in the genetic code?

Hence, this paper is an attempt to make the first step toward the search of correct answers to the above questions, if not completely, then to present other possible arithmetical regularities, as well as other possible classifications-systematizations of AAs, which would take us, *per se*, at least closer to an adequate response.

## 2. PRELIMINARIES

We start from Shcherbak's basic Table (here: Table 1)[1], given here with our shading, which indicates that the determination of the number of nucleons involves only one column (the seventh) and only one row [a part of the first row: the numbers with the same digits (111, 222 and 333)]; in other words, the determination involves only the digital patterns of the nucleon number notations that are mediated by the Pythagorean triplet, one and only existed within natural numbers series: 3-4-5.

[1] Cf. Table 1 in this paper with Shcherbak's Tables 1 and 2, in relation to Figure 1 in his source work (Shcherbak, 1994).

In fact, only the number of nucleons in four-codon AAs ("Group IV" in Shcherbak's labelling) is directly determined by the Pythagorean triplet: $3^2$ x 037 = 333, $4^2$ x 037 = 592, $5^2$ x 037 = 925. However, the fact that the numerical notation "333" participates in this Pythagorean determination is sufficient reason to conclude that the determination of the number of nucleons in the non-four-codon AAs ("quasi group III-II-I" in Shcherbak's labelling) is also a Pythagorean determination; albeit indirect, through a connection between notations "111", "222" and "333" (cf. Figure 1 in Shcherbak, 1994).

| 1 | 2 | 3 | 4 | 5 | 6 | 7 | 8 | 9 |
|---|---|---|---|---|---|---|---|---|
| 037 | 074 | 111 | 148 | 185 | 222 | 259 | 296 | 333 |
| 370 | 407 | 444 | 481 | 518 | 555 | 592 | 629 | 666 |
| 703 | 740 | 777 | 814 | 851 | 888 | 925 | 962 | 999 |

| | |
|---|---|
| 111 | 259 |
| 222 | 592 |
| 333 | 925 |

**Table 1.** Shcherbak's Table of multiples of number 037, a little simplified and with our appendix on the right. (Table 1 in Shcherbak, 1994.) For details see the text.

When it comes to the Pythagorean triplet, in the said context, it is important to know that the Pythagorean triangle 5-4-3 is actually the first in the series of the so-called Diophantus' triangles (not counting the zeroth triangle, which is not really a triangle because it lacks one straight line). It should also be noted that one cathetus of any Diophantus' triangles belongs to the series of odd integers, and the other is generated from the series whose first term is zero, and the following members are respectively increased by 4n (n = 1, 2, 3, 4, ...); the same applies to the hypotenuse, but it generates from the series whose initial member is number 1 (Figure A1 on the left, in Appendix A).

Finally, the Pythagorean 5-4-3 triangle belongs to the series of triangles that begins with Luca Pacioli's triangle. The series itself results from a specific generalization of the golden mean (Figure A1, on the right, in relation to Table A1, also on the right).

Having in mind all these regularities, valid for the Pythagorean triplet 3-4-5 it is difficult to accept that the determination of the number of nucleons in the genetic code, just through the Pythagorean triplet, is only a mere coincidence; or, moreover, that the referring to the Pythagorean triplet is a scholastic approach, a numerology and nothing more. On the contrary, these regularities may indicate that it makes sense *a working hypothesis (I)*, according to which in determination of the genetic code, except two inherent alphabets – 20 amino acids and four amino bases (two pyrimidines & two purines) – is involved still one "hidden alphabet", a series of natural numbers, with all its regularities and laws.

Besides starting from Shcherbak's work mentioned above, in this paper we also start from Sukhodolets' basic Table, presented 26 years ago (Sukhodolets, 1985), which demonstrated that the number of hydrogen atoms ("hydrogen" protons) in protein AAs is also accompanied by specific arithmetical regularities [Table 2 in relation to Table A2 in Appendix A; then Table 3 in relation to Tables 3.1 and 3.2, including Solutions (1) and (2) which explained in legend to Table 3.1].

$$11\underline{7} + (20 \times 4) = 19\underline{7}$$
$$08\underline{7} + (20 \times 5) = 18\underline{7} \qquad (1)$$

$$1 \rightarrow 38 = 741 \; (741 : 2 = 370.5)$$
$$1 \rightarrow \mathbf{39} = 780 \; (780 : 2 = \mathbf{39}0) \qquad (2)$$
$$1 \rightarrow 40 = 820 \; (820 : 2 = 410)$$

| n | Amino acids | Codon root | | | |
|---|---|---|---|---|---|
| | | First letter | | Second letter | |
| 5 | Gly | G | | G | |
| 7 | Ala Ser Asp cys | G U | G U | C C | A G |
| 8 | Asn | A | | A | |
| 9 | Pro Thr Glu His | C A | G C | C C | A A |
| 10 | Gln | C | | A | |
| 11 | Val Phe Met Tyr | G U | A U | U U | U A |
| 12 | Trp | U | | G | |
| 13-14 | Leu Ile Arg Lys | C A | C A | U U | G A |

**Table 2.** The Sukhodolets' Table: the number of hydrogen atoms (n) within amino acid molecules, in relation to natural numbers series: 5, (6), 7, 8, 9, 10, 11, 12, 13, 14 (Sukhodolets, 1985) (Cf. Table A2 in App. A). First letter plus second letter equals a codon root. The codon root plus third letter equals a complet codon. Within 64 codons (192 nucleotides) there are 2 x 3456 atoms; 3456 in two inner and 3456 in two outer columns of standard Genetic Code Table (GCT), all calculated as in Figures A2 and A3.

| The number of H atoms (in brackets) and conformations | | | | | | |
|---|---|---|---|---|---|---|
| G (**05**) (04) | A (**07**) (03) | S (**07**) (09) | D (**07**) (10) | C (**07**)(21) | 153 | |
| N (**08**) (16) | P (**09**)(02) | T (**09**) (08) | E (**09**) (20) | H (**09**) (24) | 298 | |
| Q (**10**) (38) | V (**11**) (08) | F (**11**) (12) | M (**11**) (20) | Y (**11**) (12) | 388 | 569/686 |
| W (**12**) (24) | R (**14**) (66) | K (**14**) (66) | I (**13**) (20) | L (**13**) (22) | 416 | Nucleon number |
| GW + AC + PH + VY + RL = 58H+ 44 nonH = 102 (622 nucl.) Out | | | | | | |
| NQ + SD + TE + FM + KI = 59H+ 43 nonH = 102 (633 nucl.) (In) | | | | | | |
| The number of conformations: | | | | | | |
| GW28 + **AC24** + PH26 + **VY20** + RL88 = 142 / **44** (Outer) | | | | | | 203 –10 |
| **NQ54** + SD19 + **TE28** + FM32 + **KI86** = **168** / 51 (Inner) | | | | | | **202 +10** |
| Dark / Light: (210 / 195) // (210 + 100) / (195 - 100) Odd / Even | | | | | | |
| Dark: (41+1) x 5 = 210; Light: (40-1) x 5 = 195 | | | | | | |

**Table 3.** The Sukhodolets' Table, with a minimal modification. Above: the system of 4 x 5 AAs. The shadow space: AAs with even number of hydrogen atoms (8, 10, 12, 14); the non-shadow space: AAs with odd number of hydrogen atoms (5, 7, 9, 11, 13). In brackets: number of hydrogen atoms – bold (after Sukhodolets, 1985) and then the number of conformations – no bold. [Number of conformations after (Popov, 1989); the calculations for the number of conformations are given in three last rows, and also in fourth, the additional row.] Nucleon number through a specific "simulation": 569 within two outer rows (as the number of neutrons, 569, in all 20 AAs: within their side chains); and 686 nucleons within two inner rows (as the number of protons, 686, in all 20 AAs: within their side chains). Nucleon number through a specific balance: 622/633 within outer/inner amino acid pairs, respectively (cf. with result 622/633 in Table 6.1; then with result 633+11 / 622 – 11 in Tables 6.1 & 7; and with result 633+10 / 622 – 10 in Fig. 1). All others as it is written.

| | | | | | | | |
|---|---|---|---|---|---|---|---|
| (a) | G (05) | A (07) | S (07) | D (07) | C (07) | 33 | 13 |
| | N (08) | P (09) | T (09) | E (09) | H (09) | 44 | 24 |
| | Q (10) | V (11) | F (11) | M (11) | Y (11) | 54 | 34 |
| | W (12) | R (14) | K (14) | I (13) | L (13) | 66 | 46 |
| | 17/18<br>**35** | 21/20<br>41 | 21/20<br>41 | 20/20<br>40 | 20/20<br>40 | **98**<br>99 | **58**<br>59 |
| | | 81/81 | | | | | |
| (b) | G (05) | A (06) | S (07) | D (09) | C (07) | 34 | 09 |
| | N (09) | P (08) | T (08) | E (10) | H (11) | 46 | 21 |
| | Q (10) | V (08) | F (12) | M (09) | Y (13) | 52 | 27 |
| | W (15) | R (12) | K (10) | I (09) | L (09) | 55 | 30 |
| | 20/19<br>39 | 18/16<br>34 | 17/20<br>37 | 18/19<br>37 | 16/24<br>40 | **98**<br>89 | **48**<br>39 |
| | | 74/74 | | | | | |
| (c) | G (05) | A (06) | S (07) | D (09) | C (07) | 88<br>99 | 38<br>49 |
| | N (09) | P (08) | T (08) | E (10) | H (11) | | |
| | Q (10) | V (08) | F (12) | M (09) | Y (13) | | |
| | W (15) | R (12) | K (10) | I (09) | L (09) | | |

**Table 3**.1. Amino acid arrangement as in modified Sukhodolets' Table (as in upper part of Table 3) with hydrogen atom number in (a) and non-hydrogen atom number in (b & c). The last column is valid for the amino acid side chains, and first to last for whole molecules. Total number of hydrogen as well as non-hydrogen atoms: 58 + 59 = 117 / 48 + 39 = 087 in side chains and 98 + 99 = 197 / 98 + 89 = 187 in whole molecules, respectively [cf. Solution (1)]. The sums of non-hydrogen atoms in two and two rows in last column of (b) are 39 and 39+9, respectively (the change for the number "9" means to make a cycle more, through module q-1 in decimal numbering system). [The uniqueness of the number 39 (through the sum from 1 to 39) is presented in Solution (2).] All other balances are self-evident.

| | | | | | | | |
|---|---|---|---|---|---|---|---|
| (a) | G (02) | A (03) | S (03) | D (04) | C (03) | **15** | **05** |
| | N (04) | P (05) | T (04) | E (05) | H (06) | 24 | 14 |
| | Q (05) | V (05) | F (09) | M (05) | Y (09) | **33** | **23** |
| | W (11) | R (06) | K (06) | I (06) | L (06) | 35 | 25 |
| | 13/09 | 09/10 | 09/13 | 10/10 | 09/15 | **48** | **28** |
| | 22 | 19 | 22 | 20 | 24 | 59 | 39 |
| | | 42/43 | | | | | |
| | | | | | | | |
| (b) | G (**02**) | A (**03**) | S (**03**) | D (**04**) | C (**03**) | | |
| | N (**04**) | P (**05**) | T (**04**) | E (**05**) | H (**06**) | **54** / 53 | |
| | Q (**05**) | V (**05**) | F (**09**) | M (**05**) | Y (**09**) | **34** / 33 | |
| | W (**11**) | R (**06**) | K (**06**) | I (**06**) | L (**06**) | | |

**Table 3**.2. The amino acid arrangement as in the modified Sukhodolets' Table (as in upper part of Table 3) with the number of carbon atoms; (a) The balances in two odd and two even rows; within whole molecules: 59 – 48 = 11 and within their side chains: 39 – 28 = 11; (b) The balances between two halves of the system; within whole molecules: 54 – 53 = 1 and within their side chains: 34 – 33 = 1.

## 3. NEW ARITHMETICAL REGULARITIES

Among the arithmetical regularities valid for number 037 in the decimal numbering system, Shcherbak has shown that this number is also present in the system of numbers which are analogs of 037 in some other numbering systems; their values for *q* (numbering system basis), going from one to another, differ by three units. Written together without specifying the numbering basis, these numbers-analogs (Shcherbaks'

"Prime Quantums", PQ) are as follows: 13, 25, 37, 49, ... As we can see, the first digit belongs to the series of natural numbers, and the second to the series of odd integers.[2]

Knowing this, it is easy to see the possibility of a specific presentation of a series of natural numbers, in the form of a matrix scheme (n x $11_q$) (Scheme 1 in Table 4 and Scheme 2 in Table 5); such a scheme along whose first diagonal appear numbers in the form of Shcherbak's numbers, analogs of number 37 (N = 13, 25, 37, 49, 5B, ...).[3]

The system of numbers in the two presented schemes (Tables 4 and 5) is a totally regulated system on several grounds; primarily by the logic of close proximity, and in accordance with two important principles: the principle of minimum change (unit change in one or both digit positions)[4] and the principle of continuity. Going from left to right, and vice versa, the change is by 01; along the second diagonal[5] by 10 (after a start with number q-1); moving vertically: by 11; and along the first diagonal (containing Shcherbaks' numbers and starting with number 01), the changes are by 12.

When it comes to changes to the vertical, then, besides being different by 11, the numbers form ordered tuples of odd and even natural numbers

---

[2] "In the close vicinity of the decimal system, for example, some number systems with the bases 4 (Quantum $13_4 = 7$), 7 ($25_7 = 19$), 10 (37), 13 ($49_{13} = 61$), etc, have the same periodic features for three-digit numbers" (Shcherbak, 1994).

[3] For N = 13, q = 1+3 = 4; for N = 25, q = 2+5 = 7; for N = 37, q = 3+7 = 10 etc. These are the real analogues, in contrast to the "simulation" analogs in the first diagonal of the arrangement of natural numbers in Tables 4 & 5. There are, in fact, the numbers that possess the form of Shcherbak's numbers, not for various, but for the one and the same numerical basis, *q*. (For example in Table 4, q = 10 and in Table 5, q = 16.) Such a simulation means a specific self-similarity at the same time.

[4] We say "in both positions" because we are here dealing with two-digit numbers, which (some of them) turn out to be the key determinants of the genetic code. This means that the "diagonal law" mentioned above is also valid for two-digit numbers, while future research is to show if there is such a law for n-digit numbers (n > 2) as well.

[5] The second diagonal contains numbers the first digits of which are generated – respectively, out of the sequence of natural numbers (0, 1, 2, 3, ...), while the second digit is number q-1.

series, alternately and periodically, after modulo $q$-1, as it is shown in Solution (3):

$$
\begin{matrix}
01 & 02 & 03 & 04 & \dots \\
03 & 04 & 05 & 06 \\
05 & 06 & 07 & 08 \\
07 & 08 & 09 & 10 \\
09 & 10 & 11 & 12 \\
11 & 12 & 13 & 14 \\
13 & 14 & 15 & 16 \\
15 & 16 & 17 & 18 \\
17 & 18 & 19 & 20 \\
\dots
\end{matrix}
\qquad (3)
$$

However, since the changes in the vertical are only by a unit in both positions ($11_q$), they are also related to the last column of the Scheme, consisting of notations $11_q$, $22_q$, $33_q$, ... and so on. In the case of a decimal numbering system, the number $11_{10}$, as such maintains the best connection to the Golden mean (Table B1 in relation to Table B2 in Appendix B).

Within decimal numbering system, as we know from Shcherbak's works (Shcherbak, 1993, 1994), the number 37 is characterized by specific and unique arithmetical regularities (Table 1). Here, however, we have shown that in Scheme 1 (Table 4) specific and unique arithmetical regularities characterize not only number 37 but also its neighbours, numbers 26 and 48, each in its own way.

## 3.1. The uniqueness of the number 26

Again we consider Scheme 1 (Table 4). If the first diagonal neighbour of the number 26, the number 16 (26 – 16 = 10) is added to number 26 and its two followers (17 and 18) are successively added to the obtained result, we get the results as in Solution (4.1):

| | | | |
|---|---|---|---|
| 26 = 26 | 26 + 42 + 59 + 77 = Y | **16** +**17** + **18** = Z | |
| 26 + **16** = 42 | Y = 204 | Z = 51 | (4.1) |
| 42 + **17** = 59 | | Z = Y/4 | |
| 59 + **18** = 77 | Y/4 = 51 | | |

With three addings (16 +17 +18 = 51 = Z) we obtained three new results, and with the inclusion of the initial number 26 – four results. Their sum is 204 (26 + 42 + 59 + 77 = 204 = Y = 4Z), exactly four times greater than the sum of the three addings (16+17+18 = 51 = Z). But this connection of two equalities is a single and unique case in the entire system of numbers within Scheme 1 (Table 4), in other words within the set of natural numbers. Moreover, it appears that this is the zeroth case in the 4$^{th}$ column of Scheme 1 in the decimal numbering system (Table C1 in Appendix C), and also the zeroth case within all Shcherbak's numbering systems (q = 4, 7, **10**, 13, 16, … ) (cf. Table C2).

### 3.1.1. The uniqueness of number 26 through unique pair 25-36

The uniqueness of the number 26 is expressed not only through the difference 26 - 10 = 16, but also through the sum 26 + 10 = 36, where number 36, as the second diagonal neighbour of the number 26, appears to be the member of a unique pair 25-36; unique case in the entire system of numbers within Scheme 1 (Table 4), and that means within the set of natural numbers. Namely, the numbers 25 and 36 are neighbours in third column of Table 4 with a difference of 11 as in all other cases, in all columns. But their square roots, as integers, appear to be also neighbours, and that is the said uniqueness [Solution (4.2) in relation to Table C3]:

$$x_1 + y_1 = 36 = 6^2 \quad (x_1 = 26;\ y_1 = 10)$$
$$x_2 + y_2 = 25 = 5^2 \quad (x_2 = 17;\ y_2 = 08)$$
$$x_1 - y_1 = 16 = 4^2$$
$$x_2 - y_2 = 09 = 3^2 \qquad (4.2)$$

## 3.2. The uniqueness of the number 48

Now we will demonstrate the uniqueness of number 48 within the system of numbers in Scheme 1 (Table 4) with one exercise: Find a number in the system of numbers of Scheme 1, in series of odd or even numbers, which with its three followers gives the same result (204) which we obtained with number 26, as it is shown above. It is immediately obvious that this is number 48, as it is shown in Solution (5):

$$48+50+52+54 = 204 \qquad (5)$$

| | | | | | | | | | | | |
|---|---|---|---|---|---|---|---|---|---|---|---|
| | ... | | | | | | | | | | |
| (-2) | ... | | | | | | | | | ... | -22 |
| (-1) | -21 | -20 | -19 | -18 | -17 | -16 | -15 | -14 | -13 | -12 | -11 |
| (0) | -10 | -09 | -08 | -07 | -06 | -05 | -04 | -03 | -02 | -01 | 00 |
| (1) | 01 | 02 | 03 | **04** | 05 | 06 | 07 | 08 | 09 | 10 | 11 |
| (2) | 12 | 13 | 14 | 15 | **16** | **17** | **18** | 19 | 20 | 21 | 22 |
| (3) | 23 | 24 | 25 | **26** | 27 | 28 | 29 | 30 | 31 | 32 | 33 |
| (4) | 34 | 35 | 36 | 37 | 38 | 39 | 40 | 41 | 42 | 43 | 44 |
| (5) | 45 | 46 | 47 | 48 | 49 | 50 | 51 | 52 | 53 | 54 | 55 |
| (6) | 56 | 57 | 58 | 59 | 60 | 61 | 62 | 63 | 64 | 65 | 66 |
| (7) | 67 | 68 | 69 | 70 | 71 | 72 | 73 | 74 | 75 | 76 | 77 |
| (8) | 78 | 79 | 80 | 81 | 82 | 83 | 84 | 85 | 86 | 87 | 88 |
| (9) | 89 | 90 | 91 | 92 | 93 | 94 | 95 | 96 | 97 | 98 | 99 |
| (A) | A0 | A1 | A2 | A3 | A4 | A5 | A6 | A7 | A8 | A9 | AA |
| (B) | B1 | B2 | B3 | B4 | B5 | B6 | B7 | B8 | B9 | BA | BB |

**Table 4** (Scheme 1). A specific arrangement of natural numbers in decimal numbering system with Shcherbak's "simulation" analogs (13, 25, 37, 49 ...) within the first diagonal (cf. footnote 3). [The numbers 3, 4 and 5 appear to be unique through Pythagorean law; the uniqueness of the number 78 one can see in Solution (2) and legend to Table 3.1]. For all other details see the text.

It is also immediately obvious that this is the middle row in the system of numbers (in decimal number system) of Scheme 1, in Table 4. Thus, it is inevitable that every two rows at the same distance from the middle row (in analogous positions) yield twice the value of number 204 [Example: (37 + 39 + 41 + 43) + (59 + 61 + 63 + 65) = 2 x 204).] In addition, in relation to the two end, and the two central numbers of the sequence 48-50-52-54, there is a symmetrical and cross-regularity (Table C4).

| | | | | | | | | | | | | | | | | | |
|---|---|---|---|---|---|---|---|---|---|---|---|---|---|---|---|---|---|
| (1) | 01 | 02 | 03 | 04 | 05 | 06 | 07 | 08 | 09 | 0A | 0B | 0C | 0D | 0E | 0F | 10 | 11 |
| (2) | 12 | 13 | 14 | 15 | 16 | 17 | 18 | 19 | 1A | 1B | 1C | 1D | 1E | 1F | 20 | 21 | 22 |
| (3) | 23 | 24 | 25 | 26 | 27 | 28 | 29 | 2A | 2B | 2C | 2D | 2E | 2F | 30 | 31 | 32 | 33 |
| (4) | 34 | 35 | 36 | 37 | 38 | 39 | **3A** | **3B** | **3C** | 3D | 3E | 3F | 40 | 41 | 42 | 43 | 44 |
| (5) | 45 | 46 | 47 | 48 | 49 | **4A** | 4B | 4C | 4D | 4E | 4F | 50 | 51 | 52 | 53 | 54 | 55 |
| (6) | 56 | 57 | 58 | 59 | 5A | 5B | 5C | 5D | 5E | 5F | 60 | 61 | 62 | 63 | 64 | 65 | 66 |
| (7) | 67 | 68 | 69 | 6A | 6B | 6C | 6D | 6E | 6F | 70 | 71 | 72 | 73 | 74 | 75 | 76 | 77 |
| (8) | 78 | 79 | 7A | 7B | 7C | 7D | 7E | 7F | 80 | 81 | 82 | 83 | 84 | 85 | 86 | 87 | 88 |
| (9) | 89 | 8A | 8B | 8C | 8D | 8E | 8F | 90 | 91 | 92 | 93 | 94 | 95 | 96 | 97 | 98 | 99 |
| (A) | 9A | 9B | 9C | 9D | 9E | 9F | A0 | A1 | A2 | A3 | A4 | A5 | A6 | A7 | A8 | A9 | AA |
| (B) | AB | AC | AD | AE | AF | B0 | B1 | B2 | B3 | B4 | B5 | B6 | B7 | B8 | B9 | BA | BB |

**Table 5.** (Scheme 2). A specific arrangement of natural numbers in hexadecimal system with Shcherbak's "simulation" analogs (13, 25, 37, 49 ...) within first diagonal (cf. footnote 3). For details see the text.

**4. NEW CLASSIFICATIONS AND SYSTEMATIZATIONS**

## 4.1. Genetic code as an "imitation" of arithmetical regularities

Through a purely chemical analysis it is possible to find such arrangements of protein amino acids that fully correspond to the observed arithmetical regularities, related to the uniqueness of numbers 26 and 48; as if we are dealing with an "intelligent" imitation, in one possible classification and systematization.

| | | | | | | | | | |
|---|---|---|---|---|---|---|---|---|---|
| G 01(01) | S 05(31) | Y 15(107) | W 18(130) | 39 | 78 | | 269 | 518 | 682 |
| A 04(15) | D 07(59) | M 11(75) | R 17(100) | 39 | | 102 | 249 | | |
| C 05(47) | T 08(45) | E 10(73) | F 14(91) | 37 | 24 **13** | | 256 | 82 x 2 **92 x 1** | +1+1-1 |
| N 08(58) | Q 11(72) | V 10(43) | I 13(57) | 42 | **89** | 102 | 230 | **481** | **573** |
| P 08(41) | H 11(81) | L 13(57) | K 15(72) | 47 | | | 251 | | |

| | | | |
|---|---|---|---|
| 26 | 42 | 59 | 77 |
| 16 | 17 | 18 | |
| (1 x 68) | | (2 x 68) | |

in: (102-01); out: (102+01)

518 = 14 x 37
481 = 13 x 37

**162 + 288 = 450**

(2 x 268) – 55 = 481 | 518

| | | |
|---|---|---|
| 682 | 573 | 55 |
| **627** | **628** | 55 |

55

**162** **288** 355 **450**
in: (633+10); out: (622-10)

**Figure 1**. A specific classification and sistematizacion of amino acids, which follow from four diversity types (Figure 2). In the shadow space there are 20 AAs with nucleon number in the brackets and atom number without the brackets, both times in molecules side chains. Regarding at the columns: 26, 42, 59 and 77 atoms, the quantums appear to be as in Solution (4.1) and Table C1. Within first two and last two columns: 1 x 68 and 2 x 68 atoms, respectively. Within two inner and two outer columns: 102 ± 1 atoms and 633 +10 & 622-10 nucleons, respectively. Regarding at the rows: there are 78 atoms within first two and 78 +11 = 89 within last two rows (about the uniqueness of the number 78 see Solution (2) and legend to Table 3.1]; within first half of the middle row 13, and within the second one 13 +11 atoms. Within two halves of shadow space (light and dark) there are also specific balances: 102 ± 00 atoms and 627/628 ± 55 nucleons (682+573 = 1255). From possible six permutations: **268**, 286, **628**, **682**, 826, 862, the three are in the "game" (bold and underlined) and three are not; the sums of the "chosen" three [(268+628+682 = 2 x 789); (286+826+862 = 2 x 987)] correspond to nucleon number in an indirect way, as it is presented in Tables D3 and D4, in relation to Tables D1 & D2 in Appendix D. [Note: All the amino acid sequences are of the growing series from the aspect of number of atoms; all but one, in which Q-11 precedes V-10, because different stereochemical types have be distinguished: N-Q belong to alanine but V-I to valine type. (Hint: obviously, a species of principle of "uncertainty" is valid here: the order by types excludes the order by the number of atoms and vice versa).]

Figure 1 shows an arrangement of AAs (in the 4 x 5 system), with the number of atoms as in Solution (4.1) and in Table C1 (in the row starting with 26). On the other hand, in Table 6 we can see the arrangement of AAs (in the 2 x 10 system), with the number of atoms at odd and even positions as in the unique sequence, discussed above (48-50-52-54) [cf. Tables 6.1 & 6.2, and Solution (5)].

| | out | in | | out | in |
|---|---|---|---|---|---|
| | G (01) | N (08) | | G (01) | S (05) |
| | W (18) | Q (11) | | A (04) | T (08) |
| | A (04) | S (05) | | L (13) | I (13) |
| | C (05) | D (07) | | V (10) | D (07) |
| | P (08) | T (08) | | P (08) | E (10) |
| | H (11) | E (10) | | R (17) | K (15) |
| | V (10) | F (14) | | Y (15) | F (14) |
| | Y (15) | M (11) | | W (18) | Q (11) |
| | R (17) | K (15) | | H (11) | N (08) |
| | L (13) | I (13) | | C (05) | M (11) |
| Odd | 40 | 50 | | 48 | 50 |
| Even | 62 | 52 | | 54 | 52 |
| | 102 | 102 | | 102 | 102 |

**Table 6.** The outer/inner amino acid pairs which follow from Sukhodolets' Table (Table 3). On the left: the original order as in Table 3; on the right: the chemical order of singlet AAs as it is explained in Section 4.2.1. The sense and generating logic for the sequence 48-50-52-54 see in Table 6.1, and for the sequence 40-50-52-62 in Table 6.2.

| | | | | | | | | | | | |
|---|---|---|---|---|---|---|---|---|---|---|---|
| G | S | T | A | G | S | M | C | 22 | 154 | 102 / 102 | 633+11 / 622-11 |
| L | I | D | V | L | I | Q | W | 55 | 316 | | |
| P | E | K | R | P | E | K | R | 50 | 286 | | |
| Y | F | Q | W | Y | F | D | V | 46 | 300 | | |
| H | N | M | C | H | N | T | A | 31 | 199 | | |
| 48 | 50 | 52 | 54 | 48 | 50 | 52 | 54 | | | | |
| | | | | 287 | 310 | 323 | 335 | | | | |
| 102 / 102<br>622/ 633 | | | | | | | | | | | |

**Table 6.1.** The Table follows from the right subsystem in Table 6; the four non-shadow and four shadow columns of AAs follow from odd-even positions in that subsystem. In shadow space two right columns are in a vice-versa order in relation to non-shadow space. The sequence 48-50-52-54 (here as the number of atoms) is the 24$^{th}$ consecutive quad in a series of even natural numbers quads, as it is explained in the first paragraph of chapter 4.3. The balances of atom number (102:102) and nucleon number [622:633 and (622-11):633+11) are self-evident. (Note: the number of atoms and nucleons within amino acid side chains as in Figure 1.)

| | | | | |
|---|---|---|---|---|
| (1) | 10 | 20 | 22 | 32 |
| (2) | 20 | 30 | 32 | 42 |
| (3) | 30 | 40 | 42 | 52 |
| (4) | **40** | **50** | **52** | **62** |
| (5) | 50 | 60 | 62 | 72 |
| (6) | 60 | 70 | 72 | 82 |
| (7) | 70 | 80 | 82 | 92 |
| (8) | 80 | 90 | 92 | 102 |
| ... | | | | |

**Table 6.2.** The generating logic of the sequence 40-50-52-62. Knowing that this sequence must be in relation to sequence 48-50-52-54 in Table 6, it is clear that after the "start" with first possible two-digit number (the digits "0" and "1" in 10), it follows that just the sequence 40-50-52-62 must be in the middle position of a symmetrical two-digit numbers system.

## 4.2. "Outer" and "inner" amino acids in Sukhodolets' system

Let us now raise the question how – through an exact chemical analysis – we arrive at the two said arrangements of AAs, the arrangement in Figure 1, and the arrangement in Table 6? To answer this question let us go in reverse order; first – how do we get to Table 6?

In analyzing the relationship between AAs and codons, Sukhodolets started from the premise that the (standard) genetic code had to be fully completed even at the prebiotic stage (Sukhodolets, 1985). The fact that this idea is in full accordance with our hypothesis on the (prebiotic) complete genetic code (Rakočević, 2004a) was a sufficient reason for giving full faith to his arrangement (Table 2)[6], with a minimum necessary modification (Table 3)[7]; and it further meant that the analysis of Sukhodolets' system should start from several major obvious facts:

1. The positions of AAs within the system in Table 3 are strictly determined;

2. The system in Table 3 consists of two subsystems; the first subsystem: the left (small) subsystem consisting of $4^1 = 4$ AAs (G, N, Q, W)[8] in the form of four singlet sequences [(G), (N), (Q), (W)]; the second subsystem: the right (large) subsystem of $4^2 = 16$ AAs, in the form of four quaternary sequences [(ASDC), (PTEH), (VFMY), (RKIL)];

2.1. In the subsystem with 4 AAs, the outer are G and W; the inner ones N and Q;

2.2. In the subsystem with 16 AAs, alanine (A) and cysteine (C) are the outer; serine (S) and aspartic acid (D) the inner ones, and so on;

3. Besides the two subsystems referred to in the preceding paragraph, there are two sub-systems in terms of the existence of sets of even and odd numbers of hydrogen atoms in the amino acid molecules (the "even" subsystem with shading, and the "odd" subsystem, without shading).

After this analysis we can generate the left subsystem of Table 6: within the first column ("out") are the outer, and within the second one ("in") the inner AAs, given in the same order as in Sukhodolets' system in

---

[6] "В настоящей работе определенный порядок в дублетах оснований в кодонов объясняется на основании гипотезы о предсуществовании кристаллических ассоциатов из свободных молекул оснований и аминокислот" (Суходолец, 1985, с. 1589) {"In this paper, a certain order in the doublets of bases in the codons [and correspondent amino acids] is explained on the basis of the hypothesis on the pre-existence of crystalline associates of the free molecule bases and amino acids" (Sukhodolets, 1985, p. 1589)}.

[7] In the first step the exchange L, I / R, K → R, K / L, I was completed, for distinguishing AAs with even and odd numbers of hydrogen atoms; in the second step: L, I / I, L, for harmonizining the number of conformations in the column with the initial aspartic amino acid (D).

[8] It should be noted that these four AAs represent characteristic extremes: glycine (G) is the only amino acid that does not contain carbon in its side chain; tryptophan (W) is the only one out of 20 AAs with two rings; asparagine (N) and glutamine (Q) are the only two AAs which contain amide groups in their side chains.

Table 3. Now the next question makes sense: is it possible to sort AAs in both columns in accordance with their fundamental chemical nature? The answer is affirmative, as it is demonstrated in the right subsystem of Table 6 and in the next Section (in Section 4.2.1).

### 4.2.1. The chemical hierarchy of amino acids

Regarding at the Table 6 (on the right), it is self-evident that the simplest glycine (G) must be followed by a little more complex alanine (A). After alanine leucine (L) follows as its counterpart within the alanine stereochemical type, rather than valine (V), which belongs to the valine stereochemical type. Valine comes in the next step, as the first possible semicyclic amino acid. Then comes the first possible cyclic amino acid, the proline (P), as a counterpart to valine[9].

In this way we determined the order for five AAs (G, A, L, V, P), while the remaining five were the following: R, Y, W, H, C. When it is known that 18 AAs are non-sulfur ones, while only two AAs belong to the sulfur type, it is to be expected that sulfur AAs must be at the end.[10] If so, then the nitrogen amino acid proline (P) is followed by another nitrogen amino acid, arginine (R). Of the remaining three AAs (the forth, sulfur C, is at the end, as we concluded), it makes sense for an "ordinary" aromatic one (Y) to come first, and to be followed by two aromatic hetero-cyclic AAs; W should be the first one because it has a "normal" aromatic ring (one of two rings) and thus agrees with Y. Finally, there is a "pure" hetero-cyclic amino acid, H, in contact with sulfur amino acid C, which is "hetero"-non-cyclic.

---

[9] "The side chain of valine … follows from the shortest possible cyclic hydrocarbon, from cyclopropane, with a permanent openness and with a linkage to the ‘head’ of the amino acid through only one vertex of the cyclopropane ‘triangle’ … The proline … follows from the same source (cyclopropane), but with a permanent non-openness and with a linkage to the "head" through two vertices of the cyclopropane ‘triangle’" (Rakočević & Jokić, 1996).

[10] In relation to the set made of four kinds of atoms (C, N, O, H), the S atom features as a "hetero" atom.

Having determined the order of the outer AAs, we must also determine the sequence of the inner AAs (the rightmost column in Table 6). What is immediately obvious is the (chemically) possible pairing of singlets into doublets: L-I, R-K, Y-F and C-M. It is also obvious that the remaining pairings must be double pairings, exactly as they are presented: a simple pair of S-T with a simple pair of G-A; the non-nitrogen D-E with the non-nitrogen V- P[11]; finally, the nitrogen Q-N with the nitrogen W-H.

With this we have completely generated the system in Table 6, with strictly defined positions of AAs, such that the even and odd amino acid positions (through atom numbers) correspond to the required sequence of 48-50-52-54 atoms.

## 4.3. Toward four diversity types

Regardless of the existence of the sequence 48-50-52-54 in the system of numbers within Scheme 1 (Table 4), as well as in Sukhodolets' "In-Out" system (Table 6), it makes sense to examine this sequence in a series of natural numbers, such that it traces the "journey" of quads. It turns out to be the $24^{th}$ consecutive case [**(1) 2-4-6-8**; (2) 4-6-8-10; (3) 6-8-10-12; ...; (23) 46-48-50-52; **(24) 48-50-52-54**].[12]. In this state of affairs it makes sense to set up a *working hypothesis (II)* with the following premises:

1. If the $24^{th}$ case refers to the number of atoms in 20 AAs (in their side chains), then perhaps the first case relates precisely to the 20 amino acid molecules. That confirms that the 20 amino acid molecules can be classified into four diversity types, in terms of their chemical differences: 2, 4, 6 and 8 AAs.

[11] The nitrogen in the proline derives from the "head" of the amino acid, and only partially belongs to the side chain (cf. footnote 9).

[12] In one of the previous works (Rakočević & Jokić, 1996) we have shown that 24 amino acid singlets within Genetic Code Table appear in the form of 12 doublets (pairs) and 8 triplets, and at the same time we pointed out that the uniqueness of number 24 is manifested precisely through that distinction: "Notice that out of all doublet-triplet systems, this is the only one with two possible distinctions for doublets (i.e. six and six, and then, three and three doublets) and three possible distinctions for triplets (i.e. four and four, then two and two, and, finally, one and one triplet)" (Rakočević & Jokić, 1996).

2. The order of AAs in the 2-4-6-8 sequence must be such that the simpler of the two ones is at the beginning, while the simplest of eight AAs comes at the end.

3. It follows from the previous premise that the arrangement of 20 AAs in the four diversity types inevitably takes a linear as well as circular form, i.e. system, with strictly determined positions of AAs in it (Fig. 2).[13]

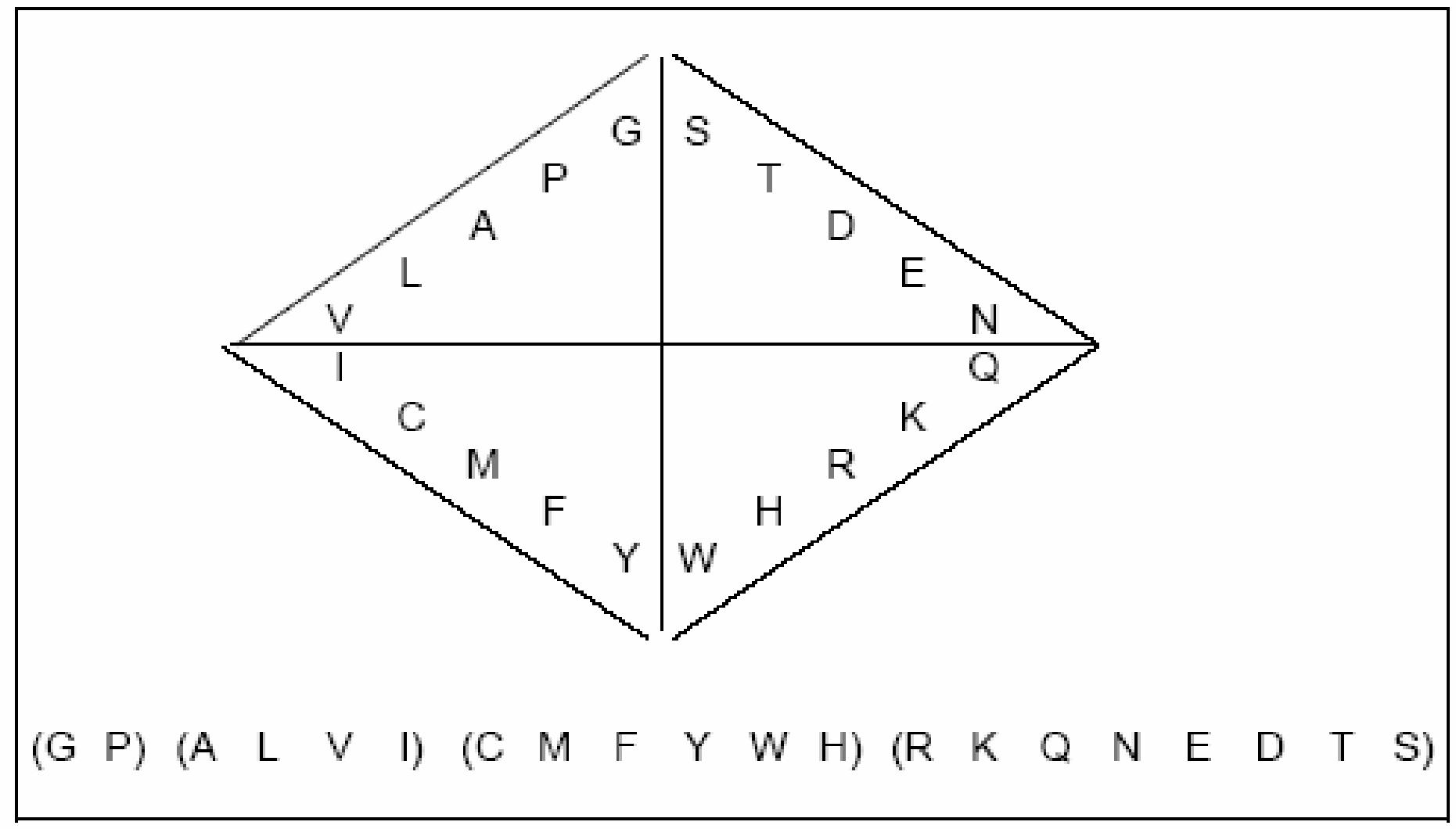

**Figure 2**. Four diversity types of protein amino acids in a linear arrangement in form of the sequence 2-4-6-8; then in form of a circular arrangement.

[13] Future research should show weather this circular arrangement is in any way related to the circular arrangements of amino acids, mediated by codons (Swanson, 1984; Castro-Chavez, 2010).

### 4.3.1. Chemical justification for the four diversity types

In classifying of 20 protein AAs into four diversity types, as shown in the linear model in Figure 2, what is immediately apparent is the splitting into 2, 4 and 8 of AA. The first type consists of two AAs (G; P), both without standard hydrocarbon side chains; the second one includes four AAs, as two pairs [(A, L), (V, I)], all with standard hydrocarbon side chains; the fourth type comprises eight AAs, as four pairs [(S, T), (D, E), (N, Q), (K, R)], all with a functional group which also exists in the amino acid functional group (wholly: $H_2N$-$CH_2$-COOH; separately: OH, COOH, $CONH_2$, $NH_2$). In addition, two AAs with the hydroxyl group in the side chain (S, T) come first, followed by two AAs with the carboxyl group in the side chain (D, E) and two of their amide derivatives (N, Q); and at the end of the set are two AAs with the amino group in the side chain (K, R).

It is a little harder to account for a set of 6 AAs, i.e. for the third diversity type, in the sense that it can represent a separate type. Following extensive analysis, the logic (and chemistry) must be as follows: the third type consists of six AAs, as three pairs [(F, Y), (H, W), (C, M)], two aromatic, two hetero aromatic and two “hetero” non-aromatic.

### 4.3.2. Four diversity versus four stereochemical types

Regarding at the Figure 2, the first diversity type is a doublet (GP) consisting from two stereochemical types [G of glycine stereochemical type, and P of proline stereochemical type]; the second one is a quadruplet (A, L, V, I); (A, L) of alanine stereochemical type and (V, I) of valine stereochemical type); then follow third diversity type (with six AAs) and fourth diversity type (with eight AAs), both of alanine stereochemical type. Precisely, because of these relationships between four diversity types and four stereochemical types, it makes sense an additional pairing between the first and second as well as third and fourth type. By this, the additional pairing between AAs of the first and second type can only participate in (G, P) and (V, I) because none of these four AAs does not belong to the alanine stereochemical type. As two new pairs appear (G,V) and (P,I), which pairs we also find in determination by two classes of aminoacyl-tRNA synthetases (Table E3 in relation to Tables E1 and E2, in Appendix E). On the other hand, in the case of an additional pairing between the third and fourth diversity types, the possibilities are even less: instead of (S, T) and (C, M) the obtained pairs are (S, C) and (T, M); such the pairs as in the determination by two classes of enzymes aminoacyl-tRNA synthetases, as it is presented in Table E3 and Solution (6). [Note: Within all pairs in two rows of Solution (6) the first member is smaller and the second one (underlined) the larger molecule. In the second (lower) row all first members are handled by the class II of enzymes aminoacyl-tRNA synthetases, while the second members – by class I.]

1 2 3 4 5 V IV III II I
**GP, [(AL), (VI)], CM, FY / WH, RK, QN, ED, TS**

**(6)**

1,3 I,4 I,4 1,3 2 II III IV V 5
**GV, SC, TM, PI, AL, DE, NQ, KR, HW, FY**

### 4.3.3. Toward an adequate amino acids arrangement

If the positions of AAs in the linear as well as the circular system, in Figure 2, are very strictly established, then it makes sense to test several possible new orders and/or arrangements of AAs (4 x 5 or 5 x 4 AAs in all of the connected sub-sets). For example, starting from the beginning we can take first four, second four, third four AAs, and so on, as it is presented in Solution (7) and, *per se*, related to our CIPS[14], presented in Solutions (8) and (9).[15]

[14] Cyclic Invariant Periodic System, arranged in cycles, starting with first cycle in middle position, as follows: 1st ST/CM, 2nd GP/VI, 3rd DE/NQ, 4th AL/KR and 5th FY/HW. [Note: The number-order in Solution (7) make the numbers taken from Solution (6), but the number-order in Solution (8) make the numbers taken directly from CIPS. But anyway, whether it comes from one or the other sequence, the same pairs of AAs appear in all three systems.] (Rakočević, 2009).

[15] As we can see, in all three systems, (6), (7) and (8), AAs appear to be in the pairs that are chemically reasonable, as we pointed out in our previous papers (Surveys 1.1 & 1.2 in Rakočević

1 2 3 4 5 V IV III II I
GP/AL, VI/CM, **FY/WH**, RK/QN, ED/TS (7)

1 3 2 IV 4 I III II 5 V
GP/VI, AL/RK, CM/TS, QN/ED, **FY/WH** (8)

2 II 4 IV 1 I III 3 5 V
GP/VI, AL/RK, CM/TS, QN/ED, **FY/WH** (9)

From the circular system in Figure 2 it is also possible to take four AAs from the central vertical line (GSYW); then, four AAs from the central horizontal line (NQVI); four from the middle point of each of the four sequences (ADRM); then, four and four from the left/right side, respectively, in relation to the middle point (PKHL/CTEF). As we can see, the five selected sequences correspond to the five sequences in Figure 1.

Each sequence in Figure 1 is arranged by the size of molecules, i.e. by the number of atoms in the side chain,[16] going from left to right; and the order of sequences is determined by the size of the first amino acid molecule in the sequence.[17] Only with such a precise and strictly regulated system can we get the desired result, the sequence 26-42-59-77, signifying the number of atoms in the four columns of AAs (in their side chains) [cf. Table 4, Solution (4.1) and Figure 1]. The obtained result also proves that our *working hypothesis (II)* is fully confirmed.

---

and Jokić, 1996; Survey 4 in Rakočević, 1998), all as the pairs in harmonic structures, presented in Appendix E.

[16] The only exception is valine, which is understandable enough when we know that valine and isoleucine belong to the same stereochemical type, the valine type. That sequence should, therefore, be understood as follows: two AAs of the alanine type (N, Q) are followed by two AAs of the valine type (V, I).

[17] The dilemma whether before N or P is resolved by the following pairs: N is followed by a smaller pair V(10) – I(13), while P by a larger one L(13) – K(15).

## 4.3.4. Toward a corresponding codons arrangement

If the *working hypothesis (II)*, related to Solution (4.1), is fully confirmed, then it makes sense to set up a *working hypothesis (III)*, related to Solution (4.2): it must be that the quantities, given in Solution (4.2), $x_1 = 26$, $y_1 = 10$ as well as $x_2 = 17$, $y_2 = 08$ in a way also contained in the genetic code. Figure 3 is an obvious and direct evidence for this. The first diversity type of AAs and corresponding **08** codons appears to be diagonally on the right within Genetic Code Table (GCT), designated here in light tones; the second one with **17** codons on the left (in dark tones); altogether in two of low-level-function diversity types there are 25 codons.

| 1st lett. | 2nd letter: U | | C | | A | | G | | 3rd letter |
|---|---|---|---|---|---|---|---|---|---|
| U | UUU | F II | UCU | | UAU | Y I | UGU | C I | U |
| | UUC | | UCC | S II | UAC | | UGC | | C |
| | UUA | | UCA | | UAA | | UGA | CT | A |
| | UUG | L I | UCG | | UAG | CT | UGG | W I | G |
| C | CUU | | CCU | | CAU | H II | CGU | | U |
| | CUC | | CCC | | CAC | | CGC | | C |
| | CUA | L I | CCA | P II | CAA | | CGA | R I | A |
| | CUG | | CCG | | CAG | Q I | CGG | | G |
| A | AUU | | ACU | | AAU | N II | AGU | S II | U |
| | AUC | I I | ACC | | AAC | | AGC | | C |
| | AUA | | ACA | T II | AAA | | AGA | | A |
| | AUG | M I | ACG | | AAG | K II | AGG | R I | G |
| G | GUU | | GCU | | GAU | D II | GGU | | U |
| | GUC | | GCC | | GAC | | GGC | | C |
| | GUA | V I | GCA | A II | GAA | | GGA | G II | A |
| | GUG | | GCG | | GAG | E I | GGG | | G |

**Figure 3**. The standard Genetic Code Table with designation of four diversity types of protein amino acids and corresponding codons: first and second type without color (in light and dark tones, respectively), but third and forth in color. The codon number: first **08**, second **17**, third **10** and fourth **26** [just as in algebraic system in Solution (4.2)]. The roman numbers designate class I and class II of aminoacyl-tRNA synthetases as in Table E3. The details see in the text.

The third diversity type of AAs and corresponding **10** codons follows, within GCT, in the next order (light blue): in column "U" *up* and *down*, and in columns "A" and "G" only *up*. The fourth type, with **26** codons (dark blue), in column "C" *up* and *down*, and in columns "A" and "G" only *down*; altogether in two of high-level-function diversity types there are 36 codons.

| | | | |
|---|---|---|---|
| G P | A L V I | C M F Y W H | R K Q N E D T S |
| 9+18=27 | 40+36 =76 | 74 + 54 = 128 | 81+ 72 = 153 |
| Atom number: (27 + 153 = **180**); (76 + 128 = **204**) | | | |
| G P **A L** V I **C M** F Y **W H** R K **Q N** E D **T S** | | | |
| Pairs on the linear arrangement in Fig. 2: Odd (633+11) / (**622 - 11**) Even | | | |
| V I **L C** A M **P F** G Y **S W** T H **D R** E K **N Q** | | | |
| Pairs on the circular arrangement in Fig. 2: Odd (569 / **686**) Even | | | |
| G **P** | A **L** V **I** | C **M** F **Y** W **H** | R **K** Q **N** E **D** T **S** |
| Singlets on the linear arrangement in Fig. 2: Odd (627-10) / (**628 + 10**) Even | | | |

**Table 7.** The relationships between four diversity types of protein amino acids, in the linear as well as circular arrangement. The calculations above: within 10 AAs of two inner types there are 180 atoms, just as within 20 amino acid "heads", i.e. 20 amino acid functional groups. On the other hand, within 10 AAs of two outer types there are 204 atoms, just as within 20 amino acid side chains. This specific "simulation" is analogue to the "simulation", valid for the number of protons and neutrons in Table 3 as well as in two last rows here. The balances within all other calculations are self-evident.

By all these results one must notice a very specific self-similarity, realized through arithmetical and/or algebraic regularities: within the system in Figure 1, related to the arithmetical system in Solution (4.1), the number 26 appears to be a minimal atom number whereas within the system in Figure 3, related to the algebraic system in Solution (4.2), the number 26 as a maximal codon number. On the other hand, since the number of atoms within codons nucleotides in two inner as well as two outer columns in GCT is 3456 both times, we can say that there is a correspondence with four exponents (3, 4, 5, 6)[18] which appear in algebraic system, given in Solution (4.2). All together, we can speak about a specific self-similarity triad (SST). [*Note*: The calculation of the number of atoms within 64 x 3 = 192 nucleotides in GCT after data: UMP 34, CMP 35, AMP 37, GMP 38; as it is presented in Figures A2 and A3.]

With these observations and, in addition, the observations marked into relationships, presented in Table 7, we conclude the analysis of relations within the four diversity types of protein amino acids and their corresponding codons.

## 5. FINAL COMMENT

As we have said in the Introduction, the intention of this paper was an attempt to answer the question why chemical classifications-systematizations of protein amino acids are followed by specific arithmetical regularities; or on the other words – what is the "physical nature" of such a phenomenon. Our answer is given in the *working hypothesis (I)*, in Preliminaries, according to which such determination follows from the fact that the genetic code is (just prebiotic) complete system, whose (harmonic) structures appear to be in a strict correspondence to the series of natural numbers, with all its regularities and laws. Evidences which are supporting *the working hypothesis (I)* we gave in the third and fourth chapter. With the knowledge that this is just

[18] In the set of all possible exponent quads [(1,2,3,4), (2,3,4,5), **(3,4,5,6)**, (4,5,6,7), ...] only two, the second and the third, contain the Pythagorean triplet. However, in the second case the number of possible "codons", 41, does not follow toward a logical Boolean cube (41"codons" + 8 stop "codons" = 49 = $7^2$). In the third case the number of possible codons, 61, follows toward a logical Boolean cube (61 codons + 3 stop codons = 64 = $8^2$). (About genetic code as a Boolean space see in Rakočević, 1997.)

so, it makes sense set out two predictions that could be useful for further researches.

**Prediction I**. The correspondence of possible classification structures of amino acids (and corresponding codons), especially the classification into four diversity types, with the series of natural numbers, should be also reflected on the entire proteome (and thus on the entire genome), measured on a representative sample of organisms.

**Prediction II**. If the first prediction be proved true, it would then mean that the statement of prediction I is also valid, *mutatis mutandis*, on the proteome and genome of individual organisms.

However, with the first prediction we actually knocking on the open door. Namely, the existing samples of proteins in which were analyzed amino acids relationships (Swanson, 1984, Taylor 1986, 1997, Kosiol et al, 2004) show that within so called Mutation ring the relationships are just such that „amino acids that are close together exchange frequently" in the evolution process (Kosiol, 2004, Figure 4 on p. 104). But now we see that interchangeable amino acids in the Mutational ring are actually these amino acid pairs that follow from the classification into four diversity type; in other words, follow from the correspondence between amino acid classification structures and the series of natural numbers [Appendix F, especially Figure F7 and Solution (F2)]. Hence, the first prediction should be understood so that it will be applied to all future representative samples of organisms.

Of course, in the above answer (to two Shcherbak's questions)[19] is hidden a new question - why is it so that the genetic code is arranged just as it is. Our answer to this "hidden" question is a *hypothesis* (for further researches) after which the solution should be related to the Mendeleyev periodic system of chemical elements (PSE). For a complete answer to this new question it is necessary, however, to resolve previously all doubts concerning to the PSE. For example, is the hydrogen really in the first and the seventh group at the same time, or only in the seventh, as we understand it (cf. Table G2 and G3, in relation to Table G1).[20]; that the PSE, in the form of short periods, appear to be a logical cube; or, in the form of long periods, a logical hypercube, as it is indicated in the original Mendeleyev (manuscript) works[21], and so on.

[19] That in connection of amino acid classification structures and arithmetical regularities appear to be the series of natural numbers.

Revealing that in determination of the genetic code, except two inherent alphabets, is involved still one "hidden alphabet" (a series of natural numbers) our intention was not only to shed light on some new classifications and new arithmetic regularities in the number of atoms and/or nucleons (which could be of use in future research in finding answers to the questions Shcherbak raised (Shcherbak, 1993, 1994)), but to provide broader answers regarding the role of numbers in the genetic code (Maddox, 1994; Dragovich & Dragovich, 2006, 2007; Dragovich, 2009; Negadi, 2009, 2011).

Apart from that, even regardless of answering these questions, it is our belief that the presented classification of protein AAs into the four diversity types will be the *key to understanding* the genetic code on a general plane as well.[22]

## Acknowledgement

I am grateful to Branko Dragovich for pointing out to me Sukhodolets' works and for a helpful, stimulating discussion about the genetic code.

---

[20] Similarly, the question of whether the third period is double or triple (the latter is also our idea; cf. Table G2 and G3).

[21] Cf. Photocopies X and XI, as well as Tables 13 and 16 in (Kedrov, 1977), that show the three-dimensionality and indicate the four-dimensionality. Cf. also photocopies III, VIII and IX in which Mendeleev demonstrated that the periodicity of elements properties are associated with the corresponding properties of their compounds (Kedrov, 1977; Rakočević, 2011: Mendeleev's Arhive) (http://www.rakocevcode.rs).

[22] Besides this, in my book (GENETIC CODE – Keys to Understanding), soon to be installed on my website (http://www.rakocevcode.rs), I will point out eight more keys to understanding the genetic code.

I would also like to thank Dragiša Janković for a wholehearted support – through many years – in a sophisticated analysis of the key elements of mathematics, especially arithmetic and number theory.

## Appendix A: Harmonic structures from original works (I)

**Box 1. Diophantus' triangles**

$$1^2 = 0^2 + 1^2$$

$$5^2 = 4^2 + 3^2$$

$$13^2 = 12^2 + 5^2$$

$$25^2 = 24^2 + 7^2$$

$$41^2 = 40^2 + 9^2$$

$$61^2 = 60^2 + 11^2$$

$$85^2 = 84^2 + 13^2$$

*Diophantus' triangles are in fact the Pythagorean triangles with first cathetus "a" taken from the od natural number series (r = 1, 3, 5, 7, 9, ...), secona cathetus "b" calculated through the "4k regularity' presented in Remark 1, within Section 3.1, and hypotenuse "c", where c = b+1.*

**Box 2. The first Luca Pacioli's triangle**

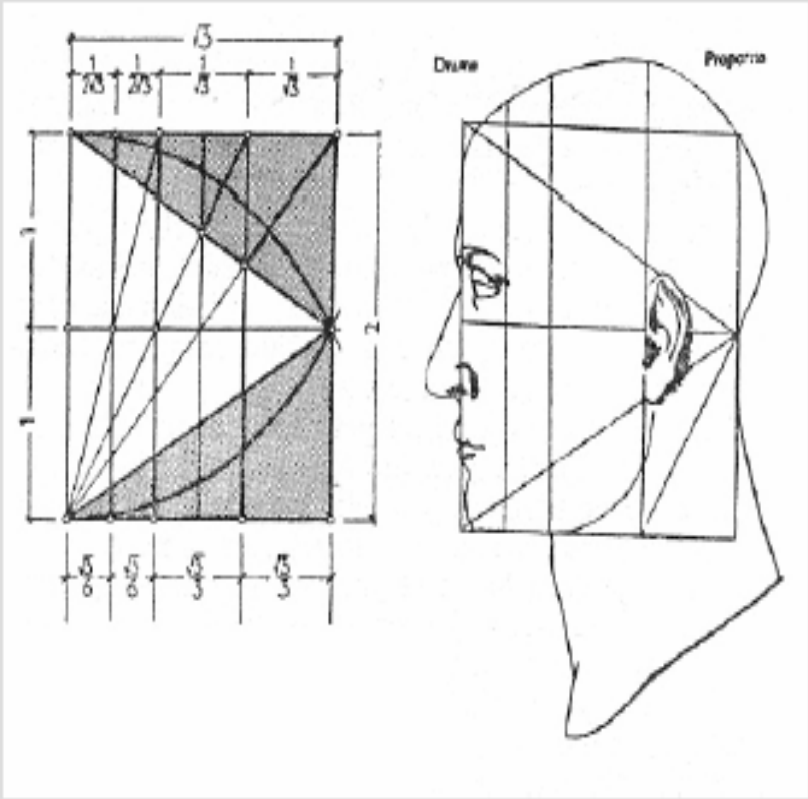

*This Luca Pacioli's triangle appears as the first triangle in the right side of Table 1. The right side of Figure follows from original Pacioli's picture, while on the left side there is a Zloković's mathematical analysis (Zloković, 1955).*

**Figure A1.** The Diophantus' triangles (on the left) and Luca Paciolis' triangle (on the right), the first triangle in the system presented in Table A1, on the right (Rakočević, 2004b).

| N | $x_1$ | | $x_2$ | | $\underline{h}$ | $m$ | $\sqrt{r}$ |
|---|---|---|---|---|---|---|---|
| 0. | $0^2$ | + | $1^2$ | = | **1** | 0 | $\sqrt{1}$ |
| | (0 | + | $1)^2$ | = | 1 | | |
| 1. | $1^2$ | + | $2^2$ | = | **5** | 4 | $\sqrt{9}$ |
| | (1 | + | $2)^2$ | = | 9 | | |
| 2. | $2^2$ | + | $3^2$ | = | **13** | 12 | $\sqrt{25}$ |
| | (2 | + | $3)^2$ | = | 25 | | |
| 3. | $3^2$ | + | $4^2$ | = | **25** | 24 | $\sqrt{49}$ |
| | (3 | + | $4)^2$ | = | 49 | | |
| 4. | $4^2$ | + | $5^2$ | = | **41** | 40 | $\sqrt{81}$ |
| | (4 | + | $5)^2$ | = | 81 | | |
| 5. | $5^2$ | + | $6^2$ | = | **61** | 60 | $\sqrt{121}$ |
| | (5 | + | $6)^2$ | = | 121 | | |
| | (...) | | | | | | |

| N | $X_1$ | | $x_2$ | | $\underline{h}$ | $M$ | $\sqrt{r}$ |
|---|---|---|---|---|---|---|---|
| 0. | $0^2$ | + | $1^2$ | = | **1** | 0 | $\sqrt{1}$ |
| | (0 | + | $1)^2$ | = | 1 | | |
| 1. | $(x_1)^2$ | + | $(x_2)^2$ | = | **2** | 1 | $\sqrt{3}$ |
| | $(x_1$ | + | $x_2)^2$ | = | 3 | | |
| 2. | $(x_1)^2$ | + | $(x_2)^2$ | = | **3** | 2 | $\sqrt{5}$ |
| | $(x_1$ | + | $x_2)^2$ | = | 5 | | |
| 3. | $(x_1)^2$ | + | $(x_2)^2$ | = | **4** | 3 | $\sqrt{7}$ |
| | $(x_1$ | + | $x_2)^2$ | = | 7 | | |
| 4. | $1^2$ | + | $2^2$ | = | **5** | 4 | $\sqrt{9}$ |
| | (1 | + | $2)^2$ | = | 9 | | |
| 5. | $(x_1)^2$ | + | $(x_2)^2$ | = | **6** | 5 | $\sqrt{11}$ |
| | $(x_1$ | + | $x_2)^2$ | = | 11 | | |
| | (...) | | | | | | |

**Table A1.** The relationships within Generalized Golden Mean, in relation to the natural numbers series (Rakočević, 2004b).

**Порядок в генетическом коде, выявляемый при разделении аминокислот на группы по числу протонов в молекуле**

| Число протонов — номер группы | Аминокислоты | Основание кодона * | |
|---|---|---|---|
| | | первое | второе |
| 5 | гли | Г | Г |
| 7 | ала сер асп цис | Г У Г У | Ц Ц А Г |
| 8 | асп | А | А |
| 9 | про тре глу гис | Ц А Г Ц | Ц Ц А А |
| 10 | глн | Ц | А |
| 11 | вал фен мет тир | Г У А У | У У У А |
| 12 | трп | У | Г |
| 13-14 | лей иле арг лиз | Ц А Ц А | У У Г А |

**Table A2.** The original Sukhodolets' Table (Sukhodolets, 1985): the numbr of hydrogen atoms within amino acid molecules, in relation to natural numbers series: 5, (6), 7, 8, 9, 10, 11, 12, 13, 14.

U C A G

12 13 | **14** | 15 16 (25)

(a)

**1 x 28** (13 + 15)

**1 x 28** (12 + 16)

(b)

| | | | |
|---|---|---|---|
| plus Ribose | 20 | | |
| | | **(2 x 6)** | |
| plus Phosphoric acid | 08 | | (26) |
| | ---- | | |
| | 28 | **(1 x 28)** | |
| minus 2 molekules $H_2O$ | 06 | **(1 x 6)** | |

UMP ( 12 + 28 ) - 6 = 34

CMP ( 13 + 28 ) - 6 = 35

AMP ( 15 + 28 ) - 6 = 37 (27)

GMP ( 16 + 28 ) - 6 = 38

----

$(2 \times 6)^2 = 144$

**Figure A2.** Atom number within four bases as well as four nucleotides (Rakočević, 1997) [Cf. legend to Table 2.]

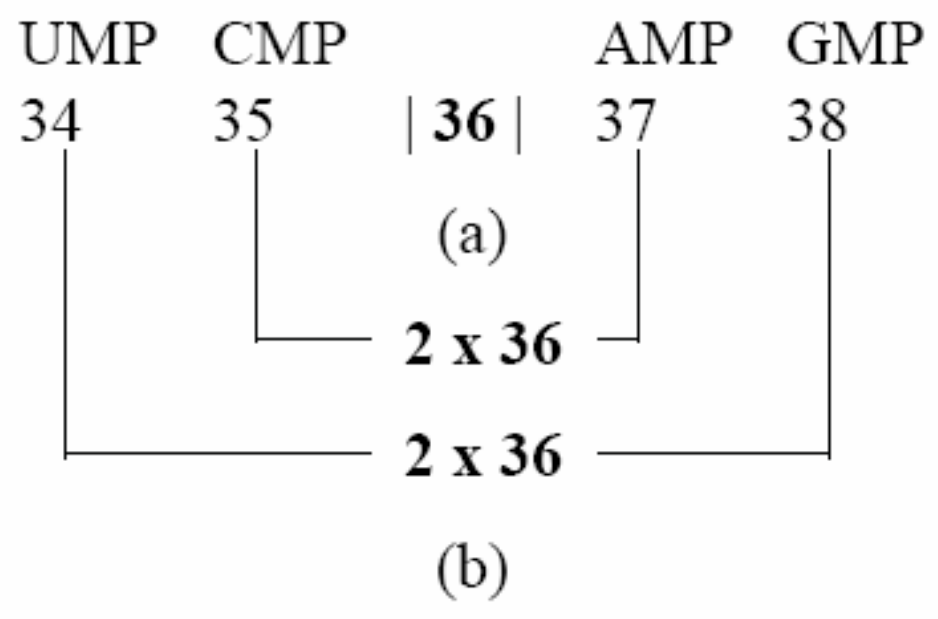

(28)

| | | | |
|---|---|---|---|
| 1/2 | x | $28^1$ | |
| 1/1 | x | $06^2$ | (a) |

-----------------------  (29)

| | | | |
|---|---|---|---|
| 1/1 | x | $28^1$ | |
| 2/1 | x | $06^2$ | (b) |

**Figure A3.** Atom number within four nucleotides (Rakočević, 1997). [As a curiosity: the numbers 6 and 28 are the first and second perfect number, respectively.]

## Appendix B: Golden mean of natural numbers

| | a | b | a – b |
|---|---|---|---|
| 1 | 0.61803 ...($\phi^1$) | 0.38196 ...($\phi^2$) | 0.23606... ($\phi^3$) |
| 2 | 1.24 | 0.76 | 0.48 |
| 3 | 1.85 | 1.15 | 0.70 |
| ... | ... | ... | ... |
| 9 | 5.56 | 3.44 | 2.12 |
| 10 | 6.18 | 3.82 | 2.36 |
| **11** | **6.80** | **4.20** | **2.60** |
| 12 | 7.42 | 4.58 | 2.84 |
| ... | | | |

a:b = 1.618033 ...

$(1.618033 \ldots)^2 = 2.618033 \ldots$

**Table B1.** The number 11 is only and one number within the set of two-digit numbers, whose golden mean is the most "golden": through the shortest distance (0.02) between (a – b) and $(a:b)^2$.

| | a | b | a – b |
|---|---|---|---|
| **1** | 0.61803 ...($\phi^1$) | 0.38196 ...($\phi^2$) | 0.23606...($\phi^3$) |
| **11** | **6.8** (6.79837 ...) | **4.2** (4.20162 ...) | **2.6** (2.59674 ...) |
| **111** | **68.6** (68.60177 ...) | **42.4** (42.39822 ...) | **26.2** (26.20354 ...) |
| **1111** | **686.6** (686.63577 ...) | **424.4** (424.36422 ...) | **262.2** (262.27154...) |

a:b = 1.618033 ...
$(1.618033 ...)^2 = 2.618033 ...$

**Table B2.** The number 111 is more "golden" than its neighbors 11 and 1111, measured through the distance between (a – b) and [n (a : b)$^2$] (n = 1, 10, 100, respectively).

**Appendix C:** The specific relations within the system of numbers in Shemes 1 & 2 (Tables 4 & 5)

| $X_{10}$ | | I | II | III | IV | Y | Y/4 | Z | Z - Y/4 |
|---|---|---|---|---|---|---|---|---|---|
| 04 | → | 04 | -02 | -07 | -11 | -16 | -04 | -15 | -(5.5 x 2) |
| 15 | → | 15 | 20 | 26 | 33 | 094 | 23.5 | 18 | -(5.5 x 1) |
| **26** | → | **26** | **42** | **59** | **77** | **204** | **51** | **51** | **±(0.0)** |
| 37 | → | 37 | 64 | 92 | 121 | 314 | 78.5 | 84 | +(5.5 x 1) |
| 48 | → | 48 | 86 | 125 | 165 | 424 | 106 | 117 | +(5.5 x 2) |
| 59 | → | 59 | 108 | 158 | 209 | 534 | 133.5 | 150 | +(5.5 x 3) |
| 70 | → | 70 | 130 | 191 | 253 | 644 | 161 | 183 | +(5.5 x 4) |
| 81 | → | 81 | 152 | 224 | 297 | 754 | 188.5 | 216 | +(5.5 x 5) |
| . . . | | | | | | | | | |

**Table C1**. This Table follows from Table 4. The number 26, from 4th column of Table 4, appears here to be in a zeroth position; (Y = I + II + III + IV; Z = 16 + 17 + 18); (the sequence 16, 17, 18 appears to be in a diagonal neighborhood to the number 26 within Scheme 1 in Table 4). Other explanations in the text.

| q | a | $X_q$ | $X_{10}$ | $Y_{10}$ | Y/4 | $Z_{10}$ | Z - Y/4 | |
|---|---|---|---|---|---|---|---|---|
| 04 | $13_4$ | 02 | 02 | 00 | 00 | - 003 | - (3 x 1) | |
| 07 | $25_7$ | 14 | 11 | 72 | 18 | 015 | - (3 x 1) | 0 |
| **10** | **$37_{10}$** | **26** | **26** | **204** | **51** | **051** | **± 0.0** | 1 |
| 13 | $49_{13}$ | 38 | 47 | 396 | 99 | 105 | + (3 x 2) | 2 |
| 16 | $5B_{16}$ | 4A | 74 | 648 | 162 | 177 | + (3 x 5) | 3 |
| ... | | | | | | | | |

1 1 0 2 5 9 14 20 ...

0 1 2 3 4 5 6 ...

**Table C2**. This Table shows the relationships in Table 4 for decimal numbering system and analog Tables (not given here) for Shcherbak's numbering systems (q) and Shcherbak's numbers, analogs of number $37_{10}$ (a). As it is clear, decimal numbering system is a zeroth system within this arrangement, and other numbering systems (within column "q"), through their distances in relation to zeroth point, make a strict natural numbers series.

| | |
|---|---|
| $2^2 - 1^2$ | 3 |
| $3^2 - 2^2$ | 5 |
| $4^2 - 3^2$ | 7 |
| $5^2 - 4^2$ | 9 |
| $6^2 - 5^2$ | 11 |
| $7^2 - 6^2$ | 13 |
| $8^2 - 7^2$ | 15 |
| $9^2 - 8^2$ | 17 |

**Table C3**. The uniqueness of the pair 25-36 in Scheme 1 (Table 4). Details in the text.

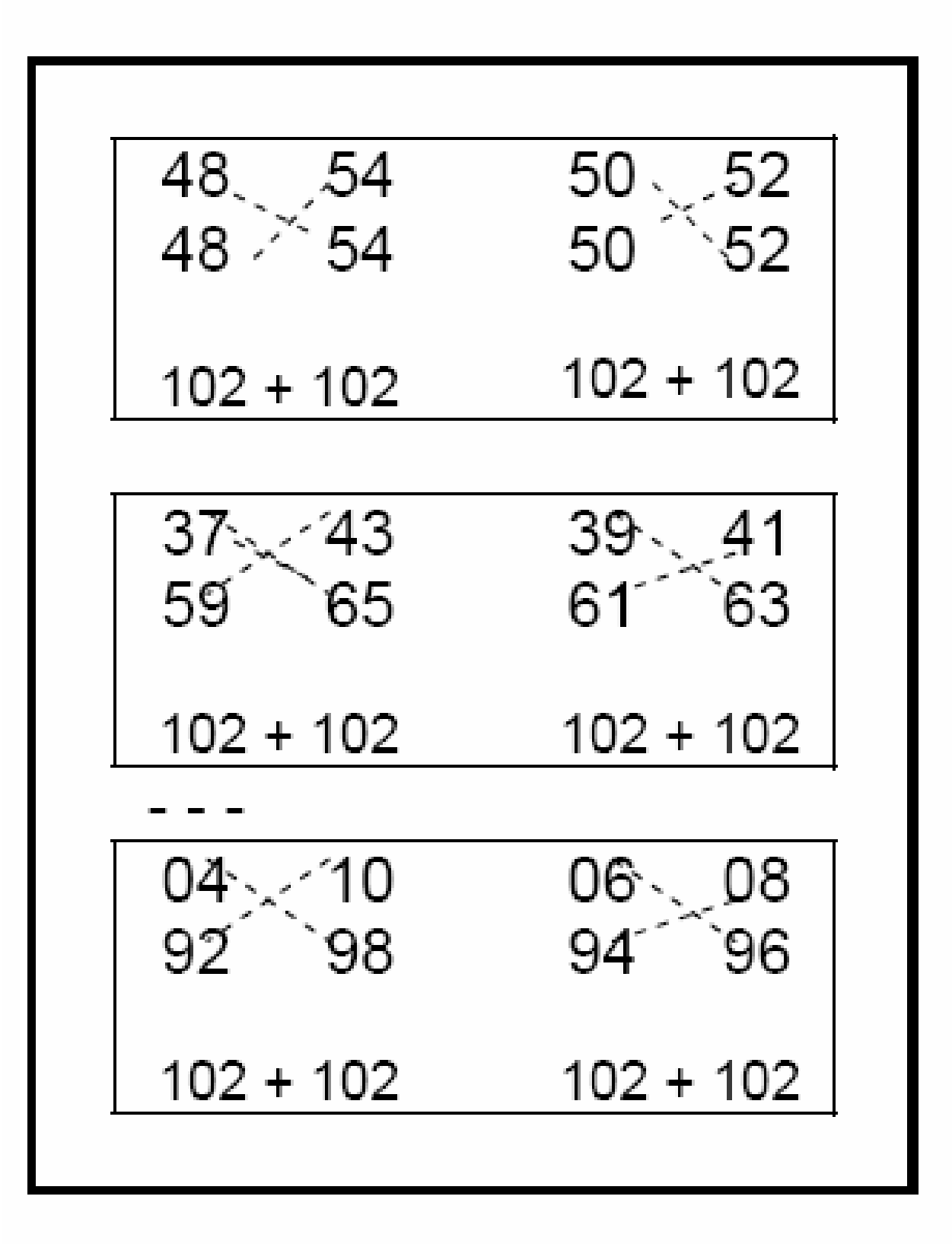

**Table C4**. From all crossing connections within Scheme 1 (Table 4) follows result 102 that we have in genetic code, as a half of all atoms in 20 AAs (their side chains).

**Appendix D:** *Relationships within standard genetic code Table*

| | | | |
|---|---|---|---|
| 1. F 91<br>2. L 57 | 1. S 31 | 6. Y 107<br>stop | 5. C 047<br>stop<br>6.W 130 |
| 2. L 57 | 2. P 41 | 7. H 81<br>8. Q 72 | 7. R 100 |
| 3. I 57<br>4.M 75 | 3. T 45 | 9. N 58<br>10. K 72 | 1. S 031<br>7. R 100 |
| 5. V 43 | 4. A 15 | 10. D 59<br>9. E 73 | 8. G 01 |
| Odd 627-1/ Even 628+1 | | | |

| | | | | | | | | | |
|---|---|---|---|---|---|---|---|---|---|
| 1.F | 2.L | 3.I | 4.M | 5.V | 6.Y | 7.H | 8.Q | 9.N | 10.K |
| 1.S | 2.P | 3.T | 4.A | 5.C | 6.W | 7.R | 8.G | 9.E | 10.D |

**Table D1**. In the standard Genetic Code Table (GCT) inevitably there are two amino acid sequences; the sequence of those AAs encoded by weaker codons (the middle base U or A) and of those AAs that are encoded by stronger codons (middle base C or G). One obtains the two "necklaces", with ten amino acids each, connected at the loop K-D (very bottom within two weak columns, U & A, there are strong codons because they possess strong bases in the first codon position). The obtained two "necklaces" are in correspondence with two strings in Table D2. Color marked amino acid pairs in correspondence with two classes of amino acids, handled by two classes of enzymes aminoacyl-tRNA synthetases (as in Table E3). It is obvious that the members within any pair are in neighborhood or near to the neighborhood.

| | | | | | | | | | | |
|---|---|---|---|---|---|---|---|---|---|---|
| Conf. N | 12 | 22 | 20 | 20 | 08 | 12 | 24 | 38 | 16 | 66 |
| Isot. N | 28 | 26 | 26 | 24 | 20 | 31 | 22 | 23 | 17 | 30 |
| PN | 49 | 33 | 33 | 41 | 25 | 57 | 43 | 39 | 31 | 41 |
| NN-1 | 91 | 57 | 57 | 75 | 43 | 107 | 81 | 72 | 58 | 72 |
| NN-T | 196 | 127 | 127 | 231 | 96 | 247 | 173 | 173 | 142 | 159 |
| M. Mass | 165.19 | 131.18 | 131.18 | 149.21 | 117.15 | 181.19 | 155.16 | 146.15 | 132.12 | 146.19 |
| AN | 14 | 13 | 13 | 11 | 10 | 15 | 11 | 11 | 08 | 15 |
| | + | + | + | + | + | − | − | − | − | − |
| | F | L | I | M | V | Y | H | Q | N | K |
| | S | P | T | A | C | W | R | G | E | D |
| | − | − | − | + | + | − | − | − | − | − |
| AN | 05 | 08 | 08 | 04 | 05 | 18 | 17 | 01 | 10 | 07 |
| M. Mass | 105.09 | 115.13 | 119.12 | 089.09 | 121.16 | 204.23 | 174.20 | 075.07 | 147.13 | 133.10 |
| NN-T | 85 | 90 | 116 | 34 | 169 | 278 | 217 | 03 | 192 | 161 |
| NN-1 | 31 | 41 | 45 | 15 | 47 | 130 | 100 | 01 | 73 | 59 |
| PN | 17 | 23 | 25 | 09 | 25 | 69 | 55 | 01 | 39 | 31 |
| Isot. N | 11 | 16 | 17 | 08 | 12 | 36 | 34 | 02 | 22 | 16 |
| Conf. N | 09 | 02 | 08 | 03 | 21 | 24 | 66 | 04 | 20 | 10 |

| | AN | M. Mass | NN-T | NN-1 | PN | Isot. N | Conf. N |
|---|---|---|---|---|---|---|---|
| Odd | 102-1 | 1369-1 | 15*1*3 | 627-1 | 343-1 | 210-1 | 203 + 1 |
| Even | 102 + 1 | 1369 + 1 | 15*0*3 | 628 + 1 | 343 + 1 | 211 + 1 | 202-1 |

**Table D2**. This table corresponds to Table 7 in Rakočević, 2004a. Plus and minus marks for non-polar and polar amino acids, respectively, according to the hydropathy index (Kyte and Doolittle, 1982). Designations are as follows; AN - number of atoms, NN1 - the number of nucleons in the first nuclide; NNT - total number of nucleons; PN - the number of protons; Isot. N - number of isotopes; Conf. N - number of conformations (Popov, 1989); M. Mass - molecular mass. The balances are selfevident.

<table>
<tr><td>F 91<br>L 57</td><td>S 31</td><td>Y 107<br>stop</td><td>C 47<br>stop<br>W 130</td><td rowspan="5">789</td></tr>
<tr><td>L 57</td><td>P 41</td><td>H 81<br>Q 72</td><td>R 100</td></tr>
<tr><td>I 57<br>M 75</td><td>T 45</td><td>N 58<br>K 72</td><td>S 31<br>R 100</td></tr>
<tr><td>V 43</td><td>A 15</td><td>D 59<br>E 73</td><td>G 01</td></tr>
<tr><td colspan="4">789</td></tr>
</table>

**Table D3**. The number of nucleons within two outer columns and two inner rows in GCT (Verkhovod, 1994). Bearing in mind nucleon number in contra-spaces (Table D4), the correspondence with natural number sequence is self-evident (**789**/987 and 456/**654**).

<table>
<tr><td>F 91<br>L 57</td><td>S 31</td><td>Y 107<br>stop</td><td>C 47<br>stop<br>W 130</td><td rowspan="5">654</td></tr>
<tr><td>L 57</td><td>P 41</td><td>H 81<br>Q 72</td><td>R 100</td></tr>
<tr><td>I 57<br>M 75</td><td>T 45</td><td>N 58<br>K 72</td><td>S 31<br>R 100</td></tr>
<tr><td>V 43</td><td>A 15</td><td>D 59<br>E 73</td><td>G 01</td></tr>
<tr><td colspan="4">654</td></tr>
</table>

**Table D4**. The number of nucleons within two outer rows and two inner columns in GCT (Verkhovod, 1994). Bearing in mind nucleon number in contra-spaces (Table D3), the correspondence with natural number sequence is self-evident (456/**654** and **789**/987).

## **Appendix E:** *Harmonic structures from original works (II)*

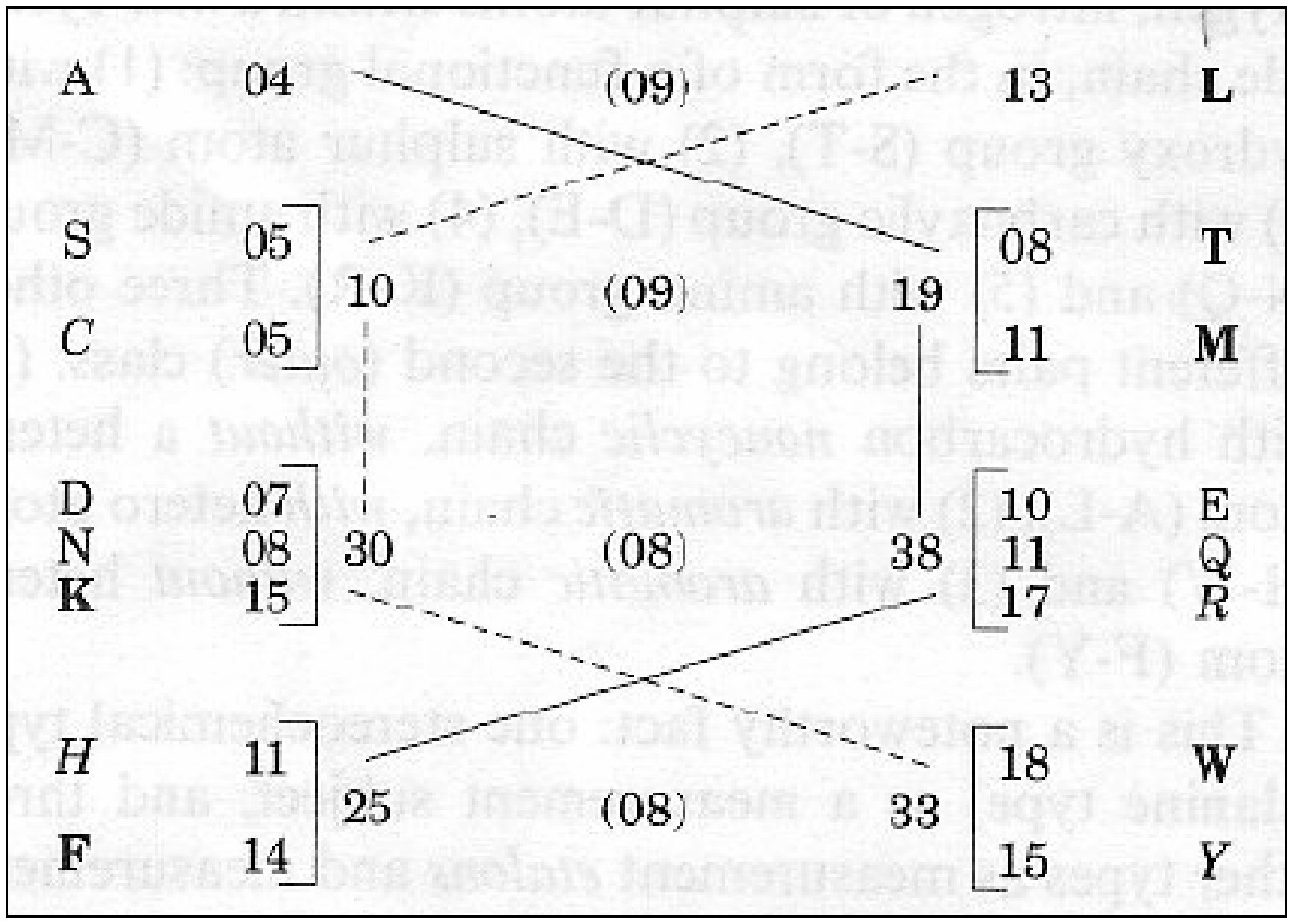

**Table E1.** This Table corresponds to Survey 1.1 in: (Rakočević & Jokić, 1996). The 16 AAs (8 pairs) of alanine stereochemical type (in further communication as the first subsystem, SubS1) are arranged in accordance to natural numbers series, as it is presented in Box 1. When this subsystem is connected with the subsystem in Table E2, then we obtain a complete overview of ten amino acid pairs; the same pairs as in upper row of Solution (e1). In the bottom row of Solution (e1) there are the pairs as in the linear arrangement of the system consisting of four diversity types, shown in Figure 2.

(GV, PI); **(**AL, ST, CM, DE, NQ, KR, HW, FY)

(e1)

(GV, PI); (AL, CM, FY / WH, RK, QN, ED, TS)

**Box E1.** *Molecule pairs hierarchy in relation to natural numbers series (I)*

**1. Aliphatic AAs (simpler than aromatic)**

1.1. Hydrocarbon AAs (start: 04 atoms);
1.2. First possible OH derivatives (start: 05 atoms);
1.3. Sulfur OH analog, i.e. SH derivatives (start: 05 atoms; S > O);
1.4. Carboxyl group derivatives (start: 07 atoms);
1.5. Amide derivatives (start: 08 atoms);
1.6. Amino derivatives (start: 15 atoms);
-------------------------------------------------------
2.2. Heteroatom derivatives (start: 11 atoms);
2.1. Aromatic hydrocarbon and its OH derivative (start: 15 atoms);

**2. Aromatic AAs (more complex than aliphatic)**

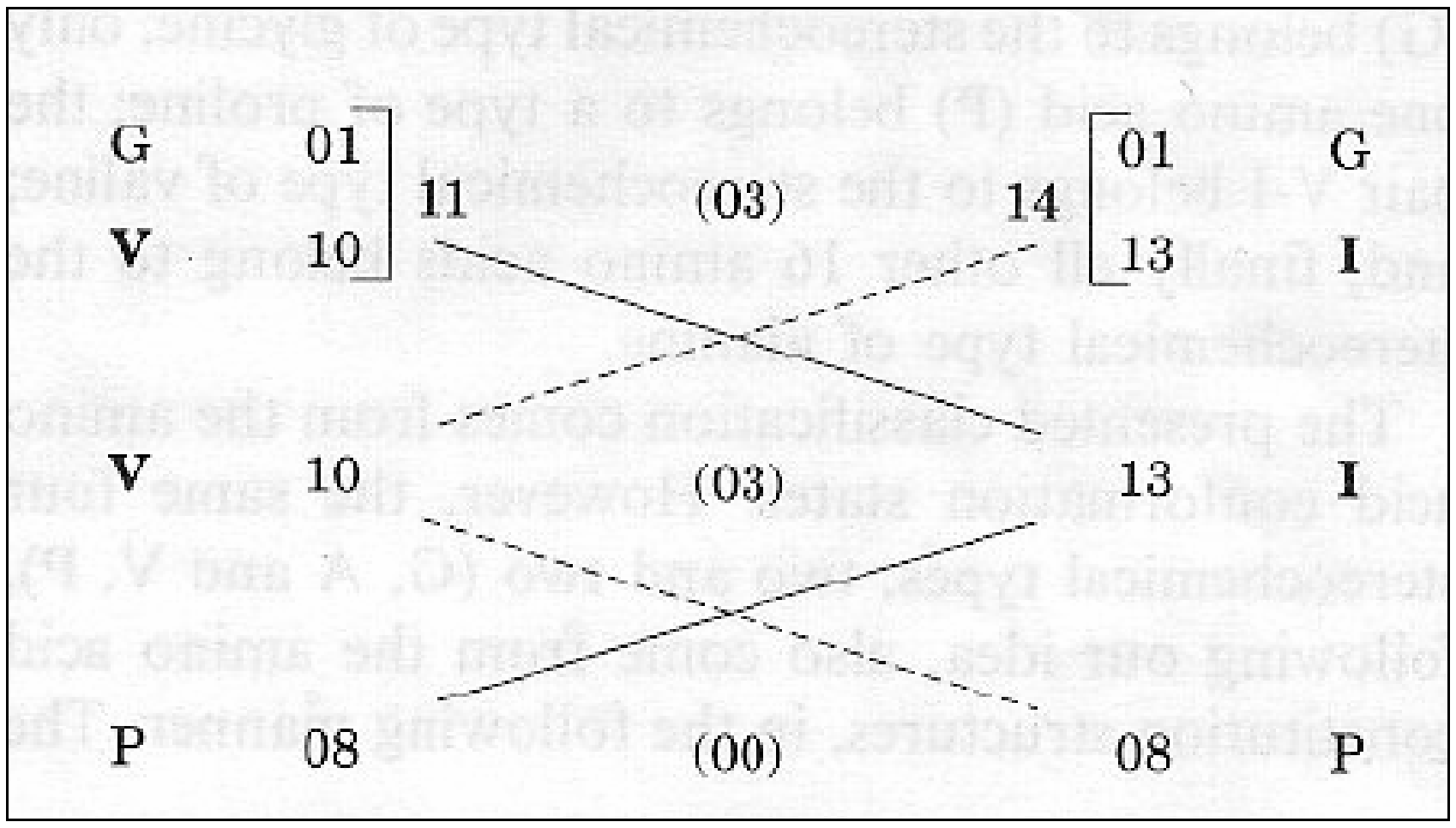

**Table E2.** This Table corresponds to Survey 1.2 in: (Rakočević & Jokić, 1996) (the second subsystem, SubS2). Here there are four AAs of non-alaninic (non-alanine) stereochemical types (G in glycinic, P in prolinic, and V-I in valinic stereochemical types) [The four stereochemical types after: (Popov, 1989; Rakočević & Jokić, 1996).] Except G-G, P-P and V-I, the possible pairs are G-V and P-I, as it is presented in the system in Table D3, in accordance to two classes of aminoacyl-tRNA synthetases. (Cf. G-P as a pair, in Fig. 2.)

**Box E2.** *Molecule pairs hierarchy in relation to natural numbers series (II)*

**1. Vertical pairs in SubS1 (Fig. D1) and SubS2 (Fig. D2)**

1.1. Glycine as the simplest (G-V) (start: 01 atom);
1.2. Serine as next (S-C) (start: 05 atoms);
1.3. Threonine as next (T-M) (start: 08 atoms);
1.4. Proline as next (P-I) (start: 08 atoms; M < I);

**2. Horizontal pairs in SubS1 (Fig. D1) and SubS2 (Fig. D2)**

2.1. Hydrocarbons (A-L) (start: 04 atoms);
2.2. Carboxyl group derivatives (D-E) (start: 07 atoms);
2.3. Amide derivatives (N-Q) (start: 08 atoms);
2.4. Amino derivatives (K-R) (start: 15 atoms);
-----------------------------------------------------
3.2. Heteroatom derivatives (H-W) (start: 11 atoms);
3.1. Aromatic hydrocarbon and its OH derivative (F-Y) (start: 15 atoms);

**3. Aromatic AAs (more complex than aliphatic)**

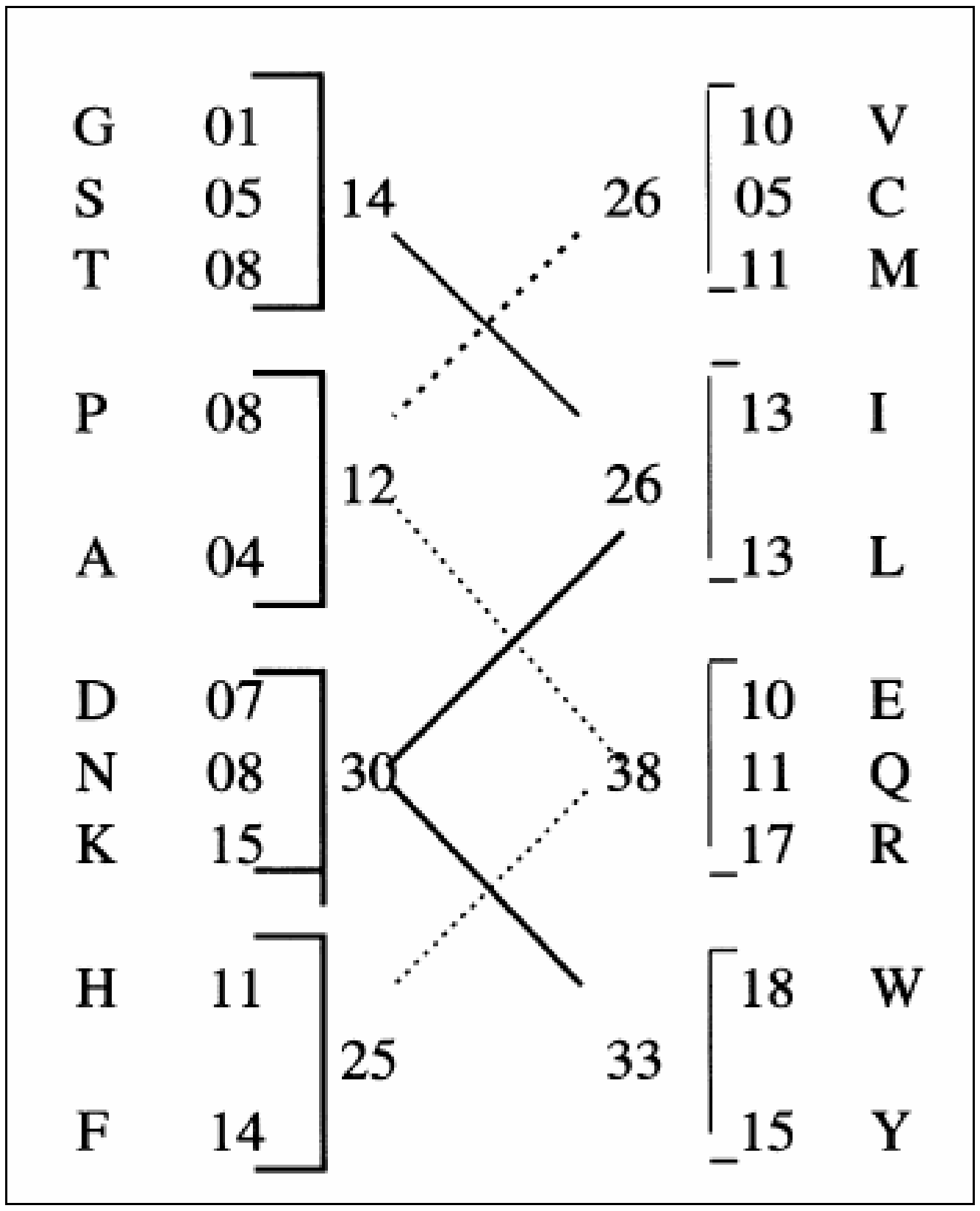

**Table E3**. This Table follows the previous two tables, the Table E1 and E2 from. When one keeps in mind the determination by two classes of aminoacyl-tRNA synthetases, it is seen that some pairs in Table E1 and Table E2 must be read as horizontal or as vertical pairs. In this Table come first vertical, then horizontal pairs, so as stated in box 2; the pairs as in upper row of Solution (e2). In the bottom row there are the pairs as in the linear arrangement of the system consisting of four diversity types, shown in Figure 2. The pairs are consistent in all but apart within the analogues ST and CM exchange occurred: the small molecule with the small, the large with the large.

**(GV, PI); (SC, TM, AL, DE, NQ, KR, HW, FY)**

(GV, PI), (AL, CM, FY / WH, RK, QN, ED, TS) (e2)

| | | | | |
|---|---|---|---|---|
| 0.0. **D** | 0.1. **N** | | 5. A | 5. L |
| 1. R | 1. F | | 4. P | 4. I |
| 2. K | 2. Y | | 3. T | 3. M |
| 3. H | 3. W | | 2. S | 2. C |
| 0.2. **E** | 0.3. **Q** | | 1. G | 1. V |

**Table E4**. This Table represents the connection between the Table E3 and Table E5. In Table E3 the molecule pair D-E make the only two charged acidic molecules. In their neighborhood are two their amide derivatives. The following six molecules, in a strict order, appear to be two triplets; the first (three basic molecules), the triplet RKH and the second one (three neither acidic nor basic molecules), the triplet FYW. Actually, the first triplet ends with the sole aromatic amino acid, which is basic charged (H), and continue with three aromatic non-electrified; and then: between the first two amino acids (D-E) comes the first triplet, and between the second two (N-Q) comes the second triplet. Thus, in such a manner is generated the left half of the system in Table E5; then comes the right half from the upper part of the system in Table E3. [Note: The correspondence with the natural numbers series is visible, on the one hand, through the sequences of two three-member chains (RKH, FYW) and two five-member chains, and on the other hand, through a specific logical (or "logical"?) square: 00 D, 01 N, 10 E, 11 Q. The chemical sense is this: the source molecule D with three derivatives; in terms of logical dimensions of the square, the first derivative is the E; then both acids generate their own amide derivatives (on the second logical dimension of square), N and Q.]

| | | | | | a | b | c | d | M |
|---|---|---|---|---|---|---|---|---|---|
| D | N | A | L | → | **189** | **189** | **221** | **221 + 3** | 485.49 = ***485*** |
| R | F | P | I | → | *289* | *289* | *341* | *341+0* | *585.70=586* |
| K | Y | T | M | → | 299 | 299 | 351 | 351 + 2 | *595.71=596* |
| H | W | S | C | → | *289* | *289* | *331* | *331+1* | *585.64=586* |
| E | Q | G | V | → | ***189*** | ***189*** | ***221*** | ***221 + 3*** | 485.50 = ***485*** |
| | | | | | 1255 | 1255 | 1465 | 1465 + 9 | 2738.04 |

**Table E5**: This table corresponds to Table 1 in our previous paper (Rakočević, 2004a); a. number of nucleons within twenty amino acid molecules (side chains), calculated after the first, i.e. the lightest nuclides (H = 1, C = 12, N = 14, O = 16, S = 32); b. the same as "a"; c. number of nucleons, calculated according to the nuclides with lowest abundance in nature (H = 2, C = 13, N = 15, O = 17, S = 36); d. number of nucleons, calculated according to the latest, i.e. heaviest nuclides (H = 2, C = 13, N = 15, O = 18, S = 36); M. molecular mass. It should be noted that for nucleon number (within all nuclides) as well as for molecule mass, two principles are valid: the principle of continuity and unit change principle (e4 - e6). That means that other possible balances of nucleon number as well as molecule mass, in accordance with a-b-c-b-a model, can not save neither the neighborhood of pair-members nor the validity of these two principles. By this, the pairs are as in upper row of Solution (e3). In the bottom row there are the pairs as in the linear arrangement of the system consisting of four diversity types, shown in Figure 2.

(GV, PI); (SC, TM, AL, DE [DN, EQ], NQ, KR, HW, FY)
(e3)
(GV, PI), (AL, CM, FY / WH, RK, QN, ED, TS)

189 – 100 – 289 – 10 – 299 – 10 – 289 – 100 – 189 (e4)

221 – 110 – 331 – 10 – 341 – 10 – 351 (e5)

485 – 101 – 586 – 10 – 596 – 101 – 586 – 101 – 485 (e6)

(a)

| | | | | | |
|---|---|---|---|---|---|
| S05 | T08 | L13 | A04 | G01 | 31 |
| D07 | E10 | M11 | C05 | P08 | 41 |
| K15 | R17 | Q11 | N08 | V10 | 61 |
| F14 | Y15 | W18 | H11 | I13 | 71 |
| 91 | | 81 | | | |

(b)

| | 91 | | 81 | |
|---|---|---|---|---|
| 29 | S05 | T08 | M11 | C05 |
| 36 | D07 | E10 | Q11 | N08 |
| 49 | K15 | R17 | L13 | A04 |
| 58 | F14 | Y15 | W18 | H11 |
| 32 | G01 | V10 | I13 | P08 |
| | 11 | | 21 | |

**Table E6**: Distribution of amino acids according to the Gaussian algorithm. According to an anecdote the small Gauss (only 9 years of age) summed up all the numbers from 1 to 100. If, however, he summed from 1 to 101, and then selected all the tens at the top to get the numbers 11, 21, 31, 41 – 61, 71, 81 and 91, then he got the patterns that are now found in the genetic code (Rakočević, 2009). The pairs as in upper row of Solution (e7). In the bottom row of Solution (e7) there are the pairs as in the linear arrangement of the system consisting of four diversity types, shown in Figure 2.

(GV, PI); (AL, CM, NQ, HW, FY, KR, DE, ST)

(e7)

(GV, PI); (AL, CM, FY / WH, RK, QN, ED, TS)

## **Appendix F:** *Harmonic structures from original works (III)*

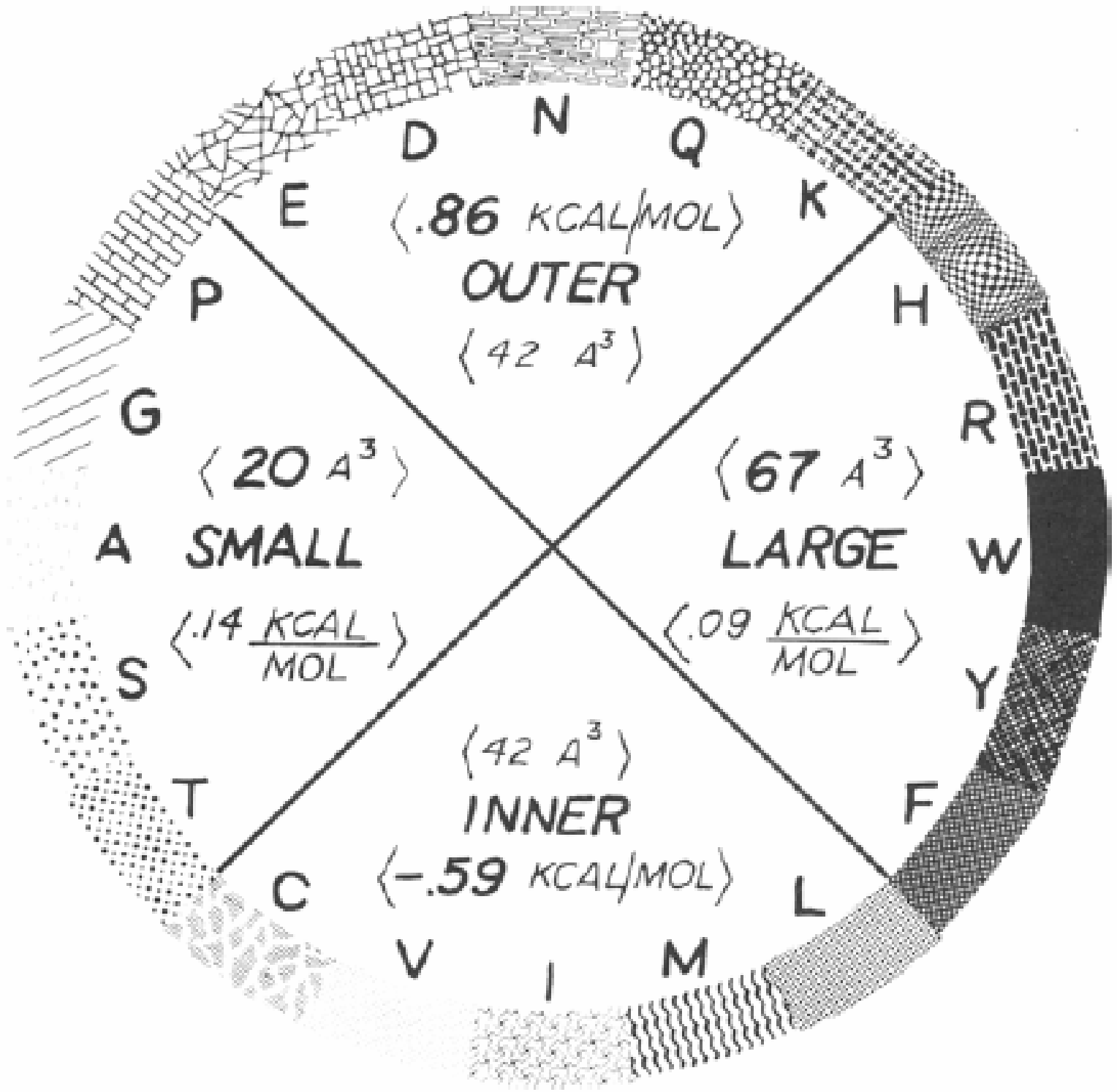

Figure 3. Mutation ring. As the codon ring expresses the minimum change relation among codons, so the mutation ring expresses the minimum change relation among the amino acids. The mutation ring shows the broader relationships among the amino acids, as well as the detailed ordering. For this the ring is quartered into the four groups, and for each group the average values of energy and volume are entered. The amino acids are marked by their one-letter codes and textured patterns. The patterns represent progressions in the physical properties of the amino acids. Dark tones are for large residues, light for small. Coarse, checkered or blotchy textures signify external residues, and smooth, delicate or even-textured patterns designate internal.

**Figure F1.** This is Figure 3 in (Swanson, 1984, p. 191)

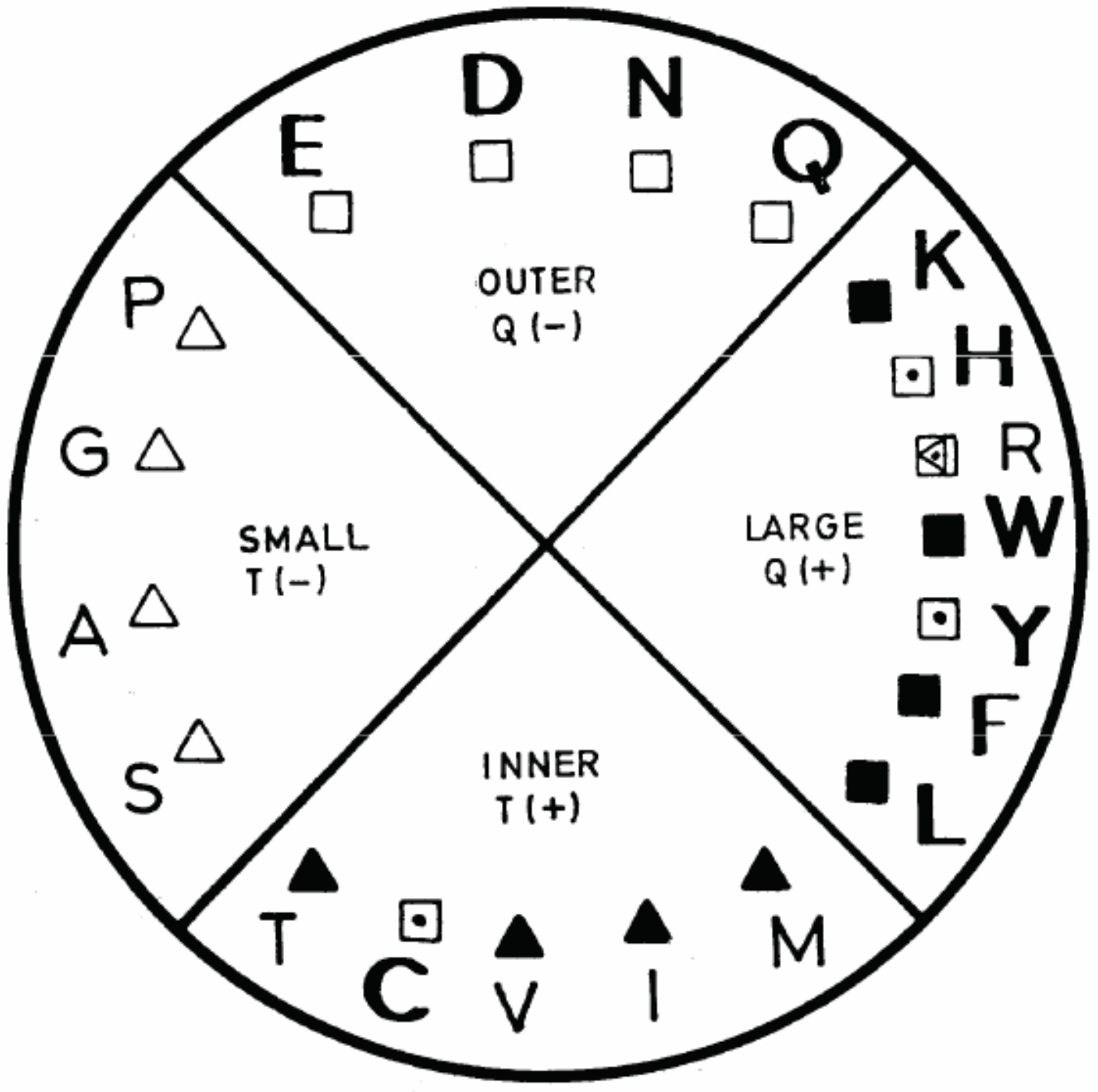

**Figure 6** *Mutation ring II. This Ring could be regarded the Mutation ring II provided that R. Swanson's Mutation Ring (Swanson, 1984, Fig. 2) is regarded the Mutation Ring I; Everything is the same as on Mutation Ring I, only the S.T.-Q.K. line is shifted by one step on both ends in relation to Mutation Ring I; and P.E.-M.L. line is shifted only on one (the other) end. The squares designate the amino acids from Space-4 and triangles designate the amino acids from Space-3. The empty squares and empty triangles designate the nonessential amino acids, otherwise they designate the essential amino acids; the dots designate the semi-essential amino acids. The lines strictly separate non-essential from yes-essential amino acids; then the lines strictly separate the Space-3 amino acids from Space-4 amino acids. There are the two exceptions: C is full-strayed; R is semi-strayed. One should note that the complementarity principle is applied as follows: outer-inner: non-essential amino acids from Space-4 are complementary with the essential amino acids from Space-3, etc.*

**Figure F2.** This is Figure 6 in (Rakočević, 1997, p. 28). Space-3 and Space-4 are from the Boolean cube; Space-3: (7) SRG, **(3) TA**, (2) IMV, (1) SP; Space-4: (6) NKDE, (5) CWR, **(4) YHQ**, (0) FL (Rakočević, 1998, Fig. 1, p. 284).

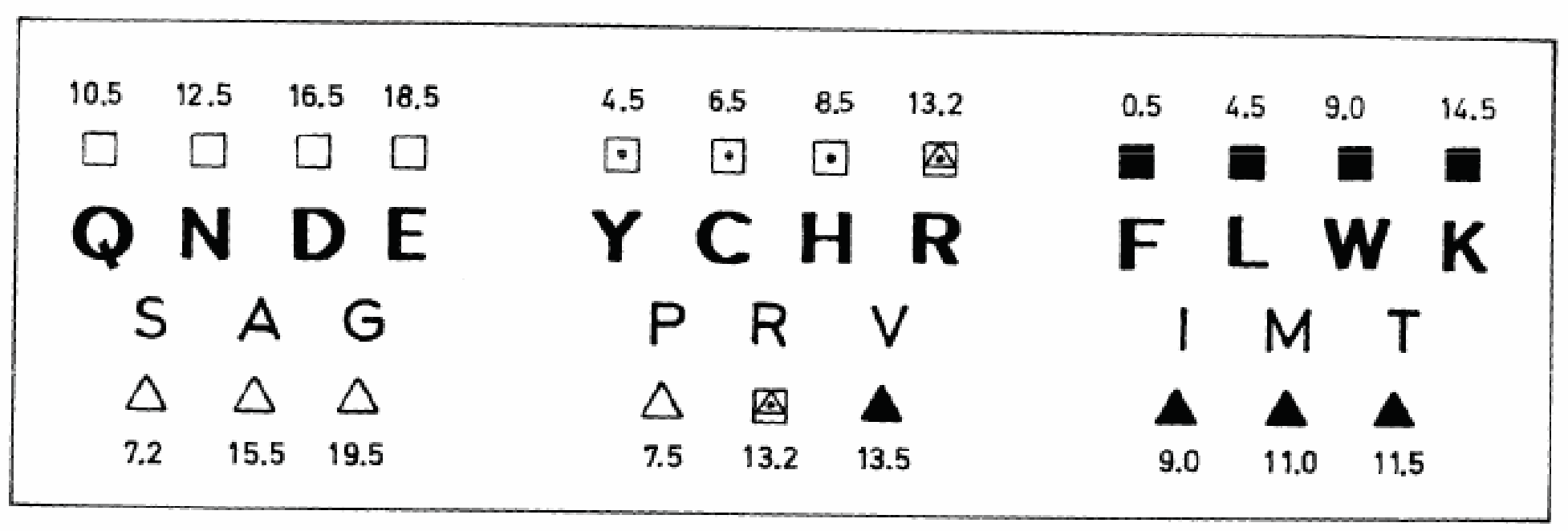

**Figure F3.** This is Figure 5 in (Rakočević, 1997, p. 27).Non-essential, semi-essential and essential protein amino acids, respectively. The order through the binary values calculated by models presented in Figures F5 and F6. The "reading" of the pairs, in relation to Mutation ring in Figure F2, is presented in Figure F4. Notice, for example the sequence QNDEPGAS**T**/VIM**T** in both Figures (F3 and F4).

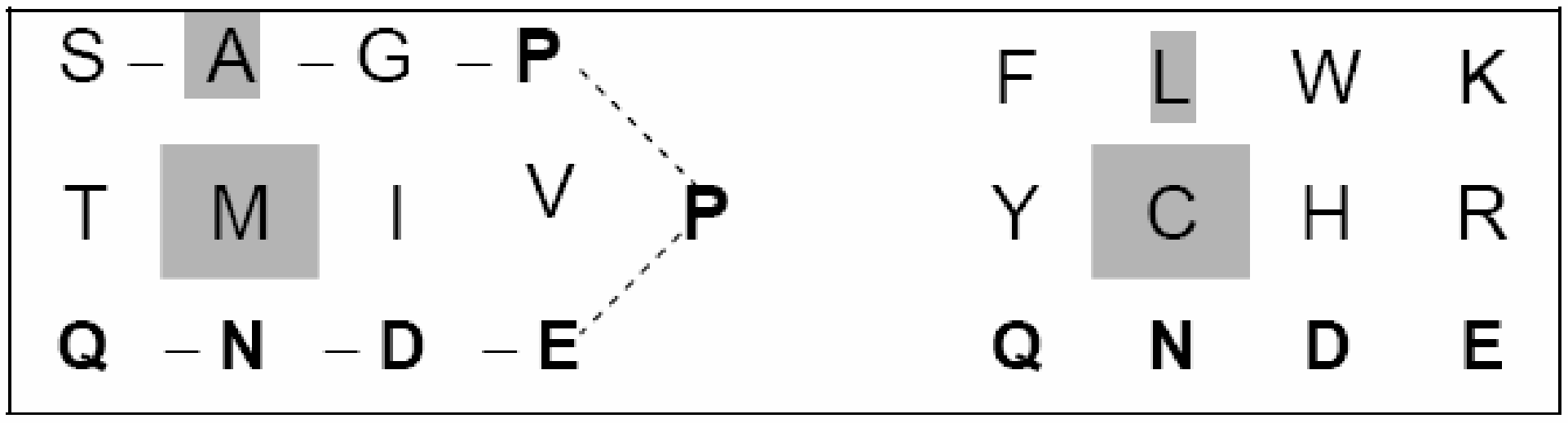

**Figure F4.** The "reading" of the pairs positions in the system presented in Figure F3. The pairs as in second (bottom) row of Solution (f1). In the first (upper) row of Solution (f1) there are the pairs as in the linear arrangement of the system consisting of four diversity types, shown in Figure 2.

1 2 3 4 5 V IV III II I

GP/AL, VI/CM, **FY/WH**, RK/QN, ED/TS

(f1)

G-P/A-L, VI/C-M, **FY/WH**, RK/QN, ED/TS

$2^0$ - third position (z) | $2^2$ $2^1$ $2^0$
4 2 1 | U = 0
$2^1$ - second position (y) | U U C = 1 | C = 1
U C U = 2 | A = 2
$2^2$ - First positon (x) | C U U = 4 | G = 3
U U A = 2 | G = 3
A C G = 13

**Figure F5.** The method of calculating the binary values of the codons (Rakočević, 1988)

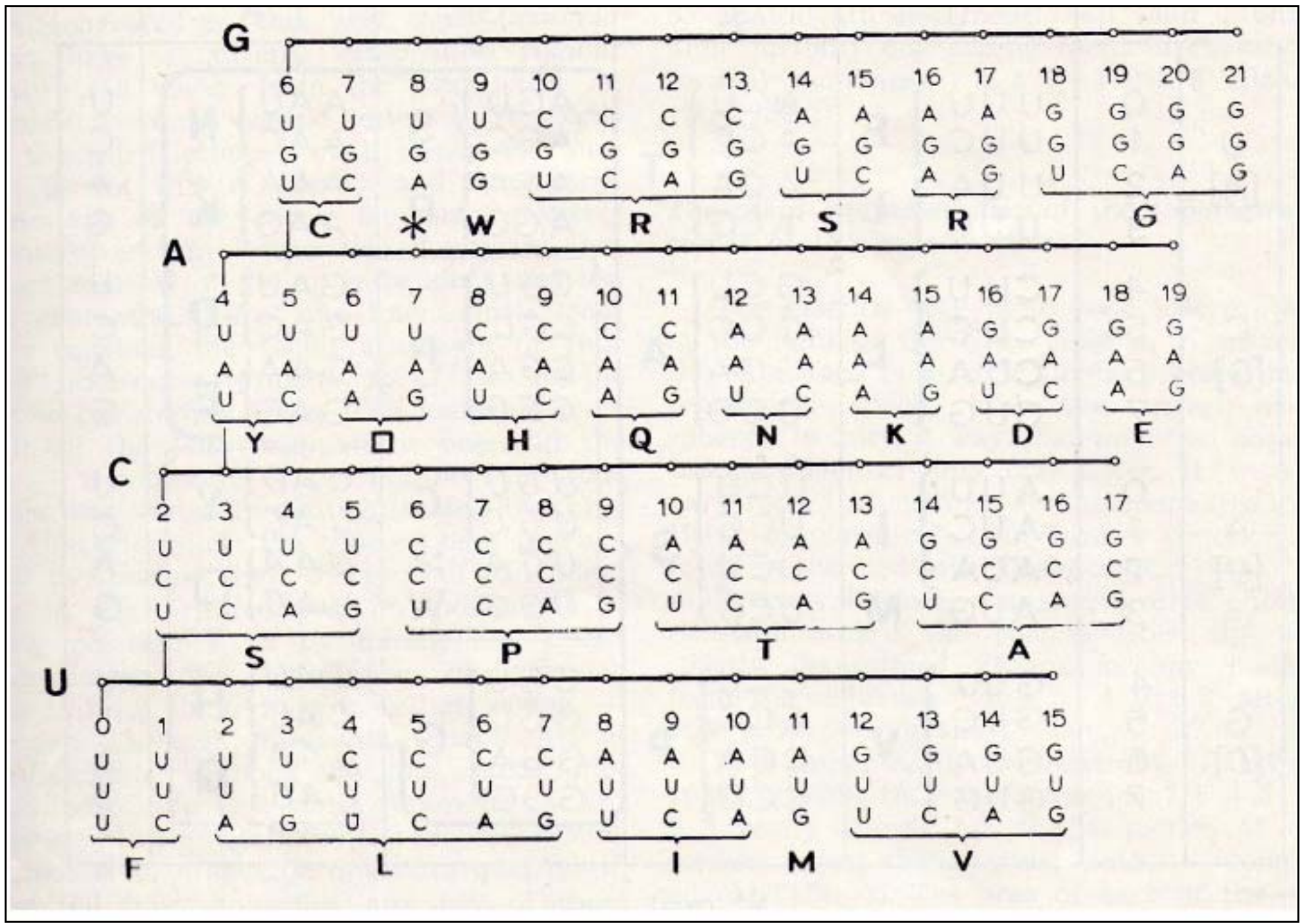

**Figure F6.** The calculated binary values of the codons (Rakočević, 1988).

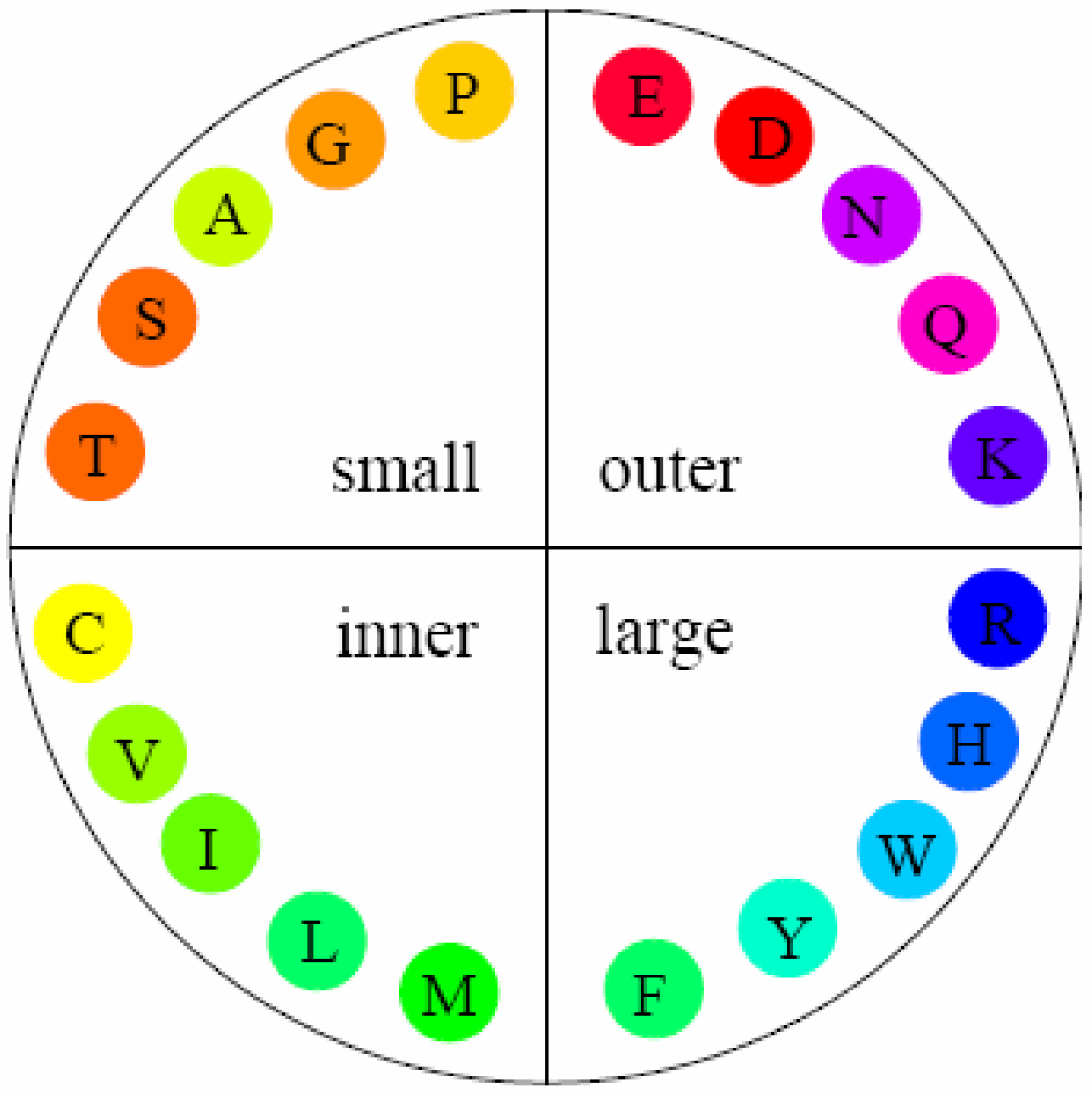

Fig. 4. Representation of the PAM matrix. This projection of the matrix by multidimensional scaling is an idealization adapted from French and Robson (1983) by Taylor (1986). The vertical axis of the circle corresponds to hydrophobicity, and consequently to whether the amino acid is mostly found in the inner or outer parts of proteins, and the horizontal axis corresponds to the molecular volume (small or large) of the amino acid. Amino acids that are close together exchange frequently. Colours used are those proposed by Taylor (1997).

**Figure F7.** This is Figure 4 in (Kosiol, 2004, p. 28). The pairs as in second (bottom) row of Solution (f2). In the first (upper) row of Solution (f2) there are the pairs as in the linear arrangement of the system consisting of four diversity types, shown in Figure 2.

1 2 3 4 5 V IV III II I
GP/AL, VI/CM, **FY/WH**, RK/QN, ED/TS

(f2)

GP/A-L, VI/C-M, **FY/WH**, RK/QN, ED/TS

## **Appendix G:** *Periodic system of chemical elements*

| | III | IV | V | VI | VII | 0 |
|---|---|---|---|---|---|---|
| | | | | | $_{1}$**H** | $_{2}$He |
| … | $_{5}$B | $_{6}$**C** | $_{7}$**N** | $_{8}$**O** | $_{9}$F | $_{10}$Ne |
| | $_{13}$Al | $_{14}$Si | $_{15}$**P** | $_{16}$**S** | $_{17}$Cl | $_{18}$Ar |
| | $_{31}$Ga | $_{32}$Ge | $_{33}$**As** | $_{34}$**Se** | $_{35}$Br | $_{36}$Kr |
| | $_{49}$In | $_{50}$Sn | $_{51}$**Sb** | $_{52}$**Te** | $_{53}$I | $_{54}$Xe |
| | … | | | | | |

**Table G1**. A part of the periodic system of chemical elements. Color indicates the possible vertically directed neighborhood from the aspects of participation in the construction of the constituents of the genetic code. [Sulfur is involved in the construction of two sulfur amino acids; under nitrogen phosphorus as builder of DNA and RNA. Blue marks a next step: selenocysteine in the set of protein amino acids; arsenic instead phosphorus in DNA and RNA of some few very specific organisms that live in extreme conditions. [“Dec. 2, 2010: NASA-supported researchers have discovered the first known microorganism on Earth able to thrive and reproduce using the toxic chemical arsenic. The microorganism, which lives in California's Mono Lake, substitutes arsenic for phosphorus in the backbone of its DNA and other cellular components.” (www.science.nasa.gov)] Finally, the color green – a hypothesis for organisms in the maximum possible extreme environments. (Note: hydrogen can not be in the first group but only in seventh, as a neighbor of helium; *see* next two tables, G2 and G3.)

| rows | periods | groups | I | II | III | IV | V | VI | VII | 0 | VIII | IX | X | XI | XII | XIII | XIV |
|---|---|---|---|---|---|---|---|---|---|---|---|---|---|---|---|---|---|
| | | | a b c | a b c | a b c | a b c | a b c | a b c | a b c | a | b c | b c | b c | c | c | c | c |
| 1 | 1 | a | | | | | | | H 1 | He 2 | | | | | | | |
| 2 | 2 | a | Li 3 | Be 4 | B 5 | C 6 | N 7 | O 8 | F 9 | Ne 10 | | | | | | | |
| 3 | 3 | a | Na 11 | Mg 12 | Al 13 | Si 14 | P 15 | S 16 | Cl 17 | Ar 18 | | | | | | | |
| 4 | 4 | a | K 19 | Ca 20 | b Sc 21 | Ti 22 | V 23 | Cr 24 | Mn 25 | | Fe 26 | Co 27 | Ni 28 | | | | |
| 5 | | b | Cu 29 | Zn 30 | a Ga 31 | Ge 32 | As 33 | Se 34 | Br 35 | Kr 36 | | | | | | | |
| 6 | 5 | a | Rb 37 | Sr 38 | b Y 39 | Zr 40 | Nb 41 | Mo 42 | Tc 43 | | Ru 44 | Rh 45 | Pd 46 | | | | |
| 7 | | b | Ag 47 | Cd 48 | a In 49 | Sn 50 | Sb 51 | Te 52 | I 53 | Xe 54 | | | | | | | |
| 8 | | a | Cs 55 | Ba 56 | b La 57 | c Ce 58 | Pr 59 | Nd 60 | Pm 61 | | Sm 62 | Eu 63 | Gd 64 | Tb 65 | Dy 66 | Ho 67 | Er 68 |
| 9 | 6 | c | Tm 69 | Yb 70 | c Lu 71 | b Hf 72 | Ta 73 | W 74 | Re 75 | | Os 76 | Ir 77 | Pt 78 | | | | |
| 10 | | b | Au 79 | Hg 80 | a Tl 81 | a Pb 82 | Bi 83 | Po 84 | At 85 | Rn 86 | | | | | | | |
| | | a | Fr 87 | Ra 88 | b Ac 89 | c Th 90 | Pa 91 | U 92 | Np 93 | | Pu 94 | Am 95 | Cm 96 | Bk 97 | Cf 98 | Es 99 | Fm 100 |
| | 7 | c | Md 101 | No 102 | a Lr 103 | a Ku 104 | Ns 105 | 106 | 107 | | 108 | 109 | 110 | | | | |
| | | b | | | | | | | | | | | | | | | |

**Table G2**. For [(s & p), d, f] elements, to the border of stability ($_{84}$Po), we have **8** times the pattern 5-3-1; then **2** times 0-3-1 and **4** times 0-0-1. All together: 9-4-1 elements. [8-4-2 and 9-4-1 as $2^{\mathbf{3}}$-$2^{\mathbf{2}}$-$2^{\mathbf{1}}$ and $\mathbf{3}^{2}$-$\mathbf{2}^{2}$-$\mathbf{1}^{2}$, respectively] The first three periods are single, each with a single row; fourth and fifth are doubles, each with two rows; sixth period is threefold, has three rows [Table 4.2 in (Rakočević, 1991), or Table 18 in (Rakočević, 1997)].

| rows | periods | groups | I a | II a | III b | IV c | V c | VI c | VII c | VIII c | IX c | X c | XI c | XII c | XIII c | XIV c | I c | II c | III c | IV b | V b | VI b | VII b | VIII b | IX b | X b | I b | II b | III a | IV a | V a | VI a | VII a | 0 a |
|---|---|---|---|---|---|---|---|---|---|---|---|---|---|---|---|---|---|---|---|---|---|---|---|---|---|---|---|---|---|---|---|---|---|---|
| 1 | 1 | a | | | | | | | | | | | | | | | | | | | | | | | | | | | | | | | H 1 | He 2 |
| 2 | 2 | a | Li 3 | Be 4 | | | | | | | | | | | | | | | | | | | | | | | | | B 5 | C 6 | N 7 | O 8 | F 9 | Ne 10 |
| 3 | 3 | a | Na 11 | Mg 12 | | | | | | | | | | | | | | | | | | | | | | | | | Al 13 | Si 14 | P 15 | S 16 | Cl 17 | Ar 18 |
| 4 | 4 | a | K 19 | Ca 20 | | | | | | | | | | | | | | | | | | | | | | | | | Ga 31 | Ge 32 | As 33 | Se 34 | Br 35 | Kr 36 |
| 5 | | b | | | Sc 21 | | | | | | | | | | | | | | | Ti 22 | V 23 | Cr 24 | Mn 25 | Fe 26 | Co 27 | Ni 28 | Cu 29 | Zn 30 | | | | | | |
| 6 | 5 | a | Rb 37 | Sr 38 | | | | | | | | | | | | | | | | | | | | | | | | | In 49 | Sn 50 | Sb 51 | Te 52 | I 53 | Xe 54 |
| 7 | | b | | | Y 39 | | | | | | | | | | | | | | | Zr 40 | Nb 41 | Mo 42 | Tc 43 | Ru 44 | Rh 45 | Pd 46 | Ag 47 | Cd 48 | | | | | | |
| 8 | 6 | a | Cs 55 | Ba 56 | | | | | | | | | | | | | | | | | | | | | | | | | Tl 81 | Pb 82 | Bi 83 | Po 84 | At 85 | Rn 86 |
| 9 | | b | | | La 57 | | | | | | | | | | | | | | | Hf 72 | Ta 73 | W 74 | Re 75 | Os 76 | Ir 77 | Pt 78 | Au 79 | Hg 80 | | | | | | |
| 10 | | c | | | | Ce 58 | Pr 59 | Nd 60 | Pm 61 | Sm 62 | Eu 63 | Gd 64 | Tb 65 | Dy 66 | Ho 67 | Er 68 | Tm 69 | Yb 70 | Lu 71 | | | | | | | | | | | | | | | |
| | 7 | a | Fr 87 | Ra 88 | | | | | | | | | | | | | | | | | | | | | | | | | 113 | 114 | 115 | 116 | 117 | 118 |
| | | b | | | Ac 89 | | | | | | | | | | | | | | | Ku 104 | Ns 105 | 106 | 107 | 108 | 109 | 110 | 111 | 112 | | | | | | |
| | | c | | | | Th 90 | Pa 91 | U 92 | Np 93 | Pu 94 | Am 95 | Cm 96 | Bk 97 | Cf 98 | Es 99 | Fm 100 | Md 101 | No 102 | Lr 103 | | | | | | | | | | | | | | | |

**Tabela G3**: Table of long periods which follows from Table G2 [Table 4.3 in (Rakočević, 1991), or Table 19 in (Rakočević, 1997)].

| | DIADS I | TRIADS I | TRIADS II | TRIADS III | DIADS II | MONAD |
|---|---|---|---|---|---|---|
| 1 | 1 H (2) VII | 2 He (2) VIII | 3 Li (2) I | 4 Be (1) II | 5 B (2) III | 6 C (2) IV |
| 2 | 7 N (2) V | 8 O (3) VI | 9 F (1) VII | 10 Ne (3) VIII | 11 Na (1) I | 12 Mg (3) II |
| 3 | 13 Al (1) III | 14 Si (3) IV | 15 P (1) V | 16 S (4) VI | 17 Cl (2) VII | 18 Ar (3) VIII |
| 4 | 19 K (3) I | 20 Ca (6) II | 21 Sc (1) III | 22 Ti (5) IV | 23 V (2) V | 24 Cr (4) VI |
| 5 | 25 Mn (1) VII | 26 Fe (4) VIII | 27 Co (1) IX | 28 Ni (5) X | 29 Cu (2) I | 30 Zn (5) II |
| 6 | 31 Ga (2) III | 32 Ge (5) IV | 33 As (1) V | 34 Se (6) VI | 35 Br (2) VII | 36 Kr (6) VIII |
| 7 | 37 Rb (2) I | 38 Sr (4) II | 39 Y (1) III | 40 Zr (5) IV | 41 Nb (1) V | 42 Mo (7) VI |
| 8 | 43 Tc (0) VII | 44 Ru (7) VIII | 45 Rh (1) IX | 46 Pd (6) X | 47 Ag (2) I | 48 Cd (8) II |
| 9 | 49 In (2) III | 50 Sn (10) IV | 51 Sb (2) V | 52 Te (8) VI | 53 I (1) VII | 54 Xe (9) VIII |
| 10 | 55 Cs (1) I | 56 Ba (7) II | 57 La (2) III | 58 Ce (4) IV | 59 Pr (1) V | 60 Nd (7) VI |
| 11 | 61 Pm (0) VII | 62 Sm (7) VIII | 63 Eu (2) IX | 64 Gd (7) X | 65 Tb (1) XI | 66 Dy (7) XII |
| 12 | 67 Ho (1) XIII | 68 Er (6) XIV | 69 Tm (1) I | 70 Yb (7) II | 71 Lu (2) III | 72 Hf (6) IV |
| 13 | 73 Ta (2) V | 74 W (5) VI | 75 Re (2) VII | 76 Os (7) VIII | 77 Ir (2) IX | 78 Pt (6) X |
| 14 | 79 Au (1) I | 80 Hg (7) II | 81 Tl (2) III | 82 Pb (4) IV | 83 Bi (1) V | 84 Po (5) VI |
| 15 | 85 At VII | 86 Rn VIII | 87 Fr I | 88 Ra II | 89 Ac III | 90 Th IV |
| 16 | 91 Pa V | 92 U VI | 93 Np VII | 94 Pu VIII | 95 Am IX | 96 Cm X |
| 17 | 97 Bk XI | 98 Cf XII | 99 Es XIII | 100 Fm XIV | 101 Md I | 102 No II |
| 18 | 103 Lr III | 104 Ku IV | 105 Ns V | 106 VI | 107 VII | 108 VIII |
| 19 | 109 IX | 110 X | 111 I | 112 II | 113 III | 114 IV |

**Table G4**. If the elements are arranged „six pro period", then it is reached the boundary of stability/instability, the 84-element, Polonium. This table is very like to the first Mendeleyev's Table which he sent to the most famous chemists of the world [the first and second photocopies in Mendeleyev's Archive, presented in Kedrov, 1977 (16 photocopies between p. 128 and p. 129)]. The classification into monads, the two groups of dyads and three of triads is determined by the number of isotopes, through the validity of two principles: the principle of continuity and minimum change principle.